\documentclass[12pt]{report}
\usepackage{latexsym,amsmath,amssymb,color}
\usepackage{ifthen}
\usepackage{verbatim}

\newtheorem{theorem}{Theorem}[section]

\def\F{I\kern-.30em{F}}
\def\P{I\kern-.30em{P}}
\def\E{I\kern-.30em{E}}
\def\build#1_#2^#3{\mathrel{\mathop{\kern 0pt#1}\limits_{#2}^{#3}}}
\def\Sum{\displaystyle\sum}

\def\dist{\mbox{\rm dist}\ }

\def\dom{\mbox{\rm dom}\ }

\def\beq{\begin{equation}}
\def\eeq{\end{equation}}
\def\ba{\begin{array}}
\def\ea{\end{array}}
\def\boa{\left.\begin{array}{llll}}
\def\eoa{\end{array}\right.}
\def\bc{\begin{center}}
\def\ec{\end{center}}

\def\qed{\ {\vrule height5pt depth0pt width5pt}\ } 

\catcode`\@=11
\newif\ifproofmode
\proofmodetrue
\def\labelcour{}

\def\@endtheorem{\hfill\ifproofmode\rlap{\tiny \kern5mm
\labelcour}\fi\endtrivlist}

\catcode`\@=12

\newenvironment{Myproof}{\textbf{Proof.} \\ }{\qed \\ }

\newcommand{\Pre}[1]{\ensuremath{\mathrm{Re} \left( #1 \right)}}

\newcommand{\demiLp}{\ensuremath{L \slash 2}}

\newcommand{\demip}{\ensuremath{1 \slash 2}}
\newcommand{\troisdemip}{\ensuremath{3 \slash 2}}
\newcommand{\Fo}{\ensuremath{\mathcal{F}}}

\newcommand{\quartp}{\ensuremath{1 \slash 4}}
\newcommand{\expo}[1]{\ensuremath{{\rm e}^{#1}}}
\newcommand{\pscal}[3]{\ensuremath{\langle #2, #3 \rangle_{#1}}}
\newcommand{\Lp}[2][2]{\ensuremath{L^{#1}(#2)}}
\newcommand{\Lps}[3][2]{\ifthenelse{\equal{#2}{0}}
                                   {\ensuremath{L^{#1}(#3)}}
                                   {\ensuremath{L^{#1}_{#2}(#3)}}}
\newcommand{\Hp}[2][1]{\ensuremath{H^{#1}(#2)}}

\newcommand{\R}{\mathbb{R}}

\newcommand{\Z}{\mathbb{Z}}

\newcommand{\N}{\mathbb{N}}

\newtheorem{lemma}{Lemma}[section]
\newtheorem{proposition}{Proposition}[section]

\newtheorem{lem:peter}{Lemma}

\newcommand{\bea}{\begin{eqnarray}}
\newcommand{\eea}{\end{eqnarray}}

\begin{document}

\begin{titlepage}

\begin{center}

{\bf Edge Currents for Quantum Hall Systems, \\
II.\ Two-Edge, Bounded and
Unbounded Geometries}

\vspace{0.1 cm}

  \setcounter{footnote}{0}
  \renewcommand{\thefootnote}{\arabic{footnote}}

{\bf Peter D.\ Hislop \footnote{Supported in part by NSF grant
      DMS-0503784.}}


  {Department of Mathematics \\
    University of Kentucky \\
    Lexington, KY 40506--0027 USA}

  \vspace{0.1 cm}

  {\bf Eric Soccorsi \footnote{also Centre de Physique Th\'eorique, Unit\'e
Mixte de Recherche 6207 du CNRS et des Universit\'es Aix-Marseille I,
Aix-Marseille II et de l'Universit\'e du Sud Toulon-Var-Laboratoire
affili\'e \`a la FRUMAM, F-13288 Marseille Cedex 9, France.}}


  {Universit\'e de la M\'editerran\'ee \\
      Luminy, Case 907 \\
  13288 Marseille, FRANCE}

\end{center}

\vspace{0.1 cm}

\begin{center}
  {\bf Abstract}
\end{center}

\noindent
Devices exhibiting the integer quantum Hall effect can be modeled
by one-electron Schr\"odinger operators describing the planar motion of an
electron in a perpendicular, constant magnetic field, and under the
influence of an electrostatic
potential. The electron motion is confined to bounded or
unbounded subsets of the
plane by confining potential barriers. The edges of the confining potential
barriers create edge currents.
This is the second of two papers in which we review recent progress
and prove explicit lower
bounds on the edge currents associated with one- and two-edge
geometries. In this paper, we study various
unbounded and bounded, two-edge
geometries with soft and hard confining potentials.
These two-edge geometries describe the electron confined to unbounded
regions in the plane, such as a strip, or to
bounded regions, such as a finite length cylinder.
We prove that the edge currents are stable under various
perturbations, provided they are suitably small relative to the magnetic
field strength, including perturbations by random potentials.
The existence of, and
the estimates on, the edge currents are independent of the spectral type of
the operator.

\end{titlepage}

\tableofcontents


\renewcommand{\thechapter}{\arabic{chapter}}
\renewcommand{\thesection}{\thechapter}

\setcounter{chapter}{1} \setcounter{equation}{0}

\section{Introduction and Main Results}

This is the second of two papers dealing with lower bound estimates on edge
currents associated with
quantum Hall devices.
The integer quantum Hall effect (IQHE) refers to the quantization of the
Hall conductivity
in integer multiples of $2 \pi e^2 / h$. The IQHE is observed in planar
quantum
devices at zero temperature
and can be described by a Fermi gas of noninteracting electrons.
This simplification reduces the study of the dynamics to
the one-electron approximation.
Typically, experimental devices consist of finitely-extended, planar
samples subject to a constant perpendicular
magnetic field $B$.
An applied electric field in the $x$-direction induces
a current in the $y$-direction,
the Hall current, and the Hall conductivity $\sigma_{xy}$ is observed to be
quantized.
Furthermore, the Hall conductivity is a function of the electron Fermi
energy, or, equivalently, the
electron filling factor, and plateaus of the Hall conductivity are
observed as the filling factor is increased.
It is now accepted that the occurrence of the plateaus is due to the
existence of localized
states near the Landau levels that are created by the random distribution
of impurities
in the sample. We refer to \cite{[HS1]} and references mentioned there for
a more
detailed discussion.
Since the earliest theoretical discussions, the existence of edge currents
has played a major role in the explanation of the quantum Hall effect.

To describe the two-edge geometries dealt with in the paper, we first
recall the
theory for the plane.
The Landau Hamiltonian $H_L(B)$ describes a particle constrained to $\R^2$,
and moving in a constant, transverse magnetic field with strength $B \geq
0$. Let $p_x = -i \partial_x$ and $p_y = - i \partial_y$ be the two
momentum operators. The operator $H_L(B)$
is defined on the dense domain $C_0^\infty (\R^2) \subset L^2 ( \R^2)$ by
\beq
H_L(B) = (-i \nabla - A)^2 = p_x^2 + (p_y - Bx)^2 ,
\eeq
in the Landau gauge for which the vector potential is $A(x,y) = B(0, x)$.
This extends to a selfadjoint operator with point spectrum given by $\{
E_n (B)  = (2n + 1)B \; | \; n =0 , 1, 2, \ldots \}$,
and each eigenvalue is infinitely degenerate.

As in \cite{[HS1]}, we define the {\it edge current} as the expectation of
the
$y$-component of
the velocity operator $V_y \equiv (p_y - Bx)$ in certain states that will
be specified below. These are states with energy concentration between two
successive Landau levels $E_n(B)$ and $E_{n+1} (B)$.

\subsection{Main Results}
Our main results in this paper can be grouped together as follows.

\begin{enumerate}
\item Two-Edge, Unbounded Geometries: We study the strip case for which the
electron is constrained to the region $-L/2 < x < L/2$, a strip of width $L
>0$. The characteristic function of the set $J$ being denoted by $\chi_J$,
>the confining potential has either one of the two forms:
\begin{enumerate}
\item Sharp Confining Potential
\beq
\label{potSharp}
V_0(x) = \mathcal{V}_0 \chi_{\{|x| > L/2 \}} (x),\ \mathcal{V}_0 > 0,
\eeq
\item Power Function Confining Potentials
\beq
\label{potp}
V_0(x)  =  \mathcal{V}_0 ( | x | - L \slash 2 )^p \chi_{ \{ |x|>L \slash 2
\} }(x),\ \mathcal{V}_0>0,\ p >1.
\eeq

\end{enumerate}
\item Two-Edge, Bounded Geometries: We study models for which the electron
on a cylinder
$C_D = \R \times D S^1$, for $D>0$, is
confined to the bounded region $[-L/2, L/2] \times DS^1$  by the sharp
confining potential (\ref{potSharp}).
\end{enumerate}
As a preamble to the investigation of these models, we shall
systematically examine the straight parabolic channel model
studied by Exner, Joye and Kovarik in \cite{[EJK]}.
In this case the confining potential is defined by
\beq
\label{potQuad}
V_0(x)  =  g^2 x^2,\ g >0,
\eeq
and it turns out this model is completely solvable,
making the estimation of the edge currents rather straightforward
in this particular case.

As in \cite{[HS1]}, we first study the edge currents for the unperturbed
Hamiltonian
$H_0 = H_L(B) + V_0$. We then examine the stability of the lower bounds
under potential perturbations.
In the sharp potential case, we prove that the lower bounds are uniform
with respect to the confining potential.
This means that we can take the limit as the size of
the confining potential becomes infinite. As a result, our results extend
to the case of Dirichlet boundary conditions along the edges.
The proof of this follows as in the first paper \cite{[HS1]}.

In all cases, the unperturbed Hamiltonian has the form
\beq
\label{unpert1}
H_0 = H_L (B) + V_0 ,
\eeq
acting on the Hilbert space $L^2 (\R^2)$.
This is a nonnegative, selfadjoint operator.
Our strategy is to analyze the unperturbed operator via the partial Fourier
transform in the $y$-variable. We write $\hat{f} (x, k)$ for this
partial Fourier transform. For the case of unbounded geometry, we have $k
\in \R
$,
whereas for the case of bounded geometry, the allowable $k$ values are
discrete.
In either case, this decomposition reduces the problem to a study of the
fibered operators of the form
\beq
\label{unpert2}
h_0(k) = p_x^2 + (k - Bx)^2 + V_0 (x) ,
\eeq
acting on $L^2(\R)$.
Since the effective,
nonnegative, potential $V_{eff} (x;k) \equiv (k - Bx)^2 + V_0 (x)$ is
unbounded as $x \rightarrow \pm \infty$,
the resolvent of $h_0(k)$ is compact and the
spectrum is discrete.
We denote the eigenvalues of $h_0(k)$
by $\omega_j (k)$, with corresponding normalized
eigenfunctions $\varphi_j (x ; k)$, so that
\beq
\label{ev1}
h_0(k) \varphi_j (x;k) = \omega_j(k) \varphi_j (x;k) ,
~~~\| \varphi_j (\cdot; k) \| = 1.
\eeq

As in \cite{[HS1]}, the properties of the curves $k \in \R \rightarrow
\omega_j
(k)$ play an
important role in the proofs. These curves are called the {\it dispersion
curves} for the unperturbed Hamiltonian (\ref{unpert1}).
The importance of the properties of the dispersion curves comes from an
application of the Feynman-Hellmann formula. To illustrate this, let us
first consider the two-edge geometry of a half-plane with the sharp
confining potential.
We note that
unlike for the case of one-edge geometries, the dispersion curves are no
longer monotonic in $k$. For simplicity, we consider in this introduction
a closed interval $\Delta_0 \subset (B , 3B)$ and a normalized
wave function $\psi$ satisfying $\psi = E_0 ( \Delta_0 ) \psi$, where
$E_0(\Delta_0)$ denotes the
spectral projection of $H_0$ associated with $\Delta_0$. Such a
function admits a decomposition of the form
\beq
\label{fourier0}
\psi (x, y) =  \frac{1}{\sqrt{2 \pi}} \int_{\omega_0^{-1} (\Delta_0)}
e^{i k y} \beta_0 (k) \varphi_0 (x ; k) ~dk,
\eeq
where the coefficient $\beta_0 (k)$ is defined by
\beq
\label{coef00}
\beta_0 (k) \equiv \langle \hat{\psi} (\cdot; k ), \varphi_0 (\cdot; k)
\rangle.
\eeq
The matrix element of the current operator $V_y$ in such a state is
\beq
\label{curr0}
\langle \psi , V_y  \psi \rangle = \int_\R ~dx \int_{\omega_0^{-1}
(\Delta_0)} ~ dk | \beta_0 (k)|^2 (k - Bx) |
\varphi_0 (x;k)|^2 .
\eeq
From (\ref{ev1}) and the Feynman-Hellmann Theorem, we find that
\beq
\label{disp0}
\omega_0 ' (k) = 2 \int_\R ~dx ~(k - Bx) ~| \varphi_0 (x;k)|^2 ,
\eeq
so that we get
\beq
\label{curr0bis}
\langle \psi , V_y  \psi \rangle = \frac{1}{2}
\int_\R  | \beta_0 (k)|^2 ~\omega_0 ' (k) ~dk.
\eeq
It follows from (\ref{curr0bis}) that in order to obtain a lower bound on
the
expectation of the current operator in the state $\psi$ we need to bound the
derivative $\omega_0 ' (k)$ from below for $k \in \omega_0^{-1} ( \Delta_0
)$.  The next step of the proof involves relating the derivative
$\omega_0 ' (k)$ to the trace of the eigenfunction $\varphi_0 (x;k)$
on the boundary of the strip.
For this, we use the formal commutator expression
\beq
\label{icommutator1}
\hat{V}_y (k) \equiv
(k-Bx) = \frac{-i}{2B} [ p_x, h_0(k)] + \frac{1}{2B} V_0' (x).
\eeq
Inserting this into the identity (\ref{disp0}), we find
\bea
\label{disp0prime}
\omega_0 ' (k) &=& 2 \langle \varphi_0 (\cdot; k ), (k-Bx) \varphi_0
(\cdot; k) \rangle \nonumber \\
&=& \frac{-i}{2B} \langle \varphi_0 (\cdot; k ),[ p_x, h_0(k)- \omega_0
(k) ]
\varphi_0(\cdot; k) \rangle + \frac{1}{B} \langle \varphi_0 (\cdot;k), V_0'
\varphi_0(\cdot; k) \rangle
         \nonumber \\
&=& \frac{\mathcal{V}_0 }{B} ( \varphi_0 (L/2;k)^2 - \varphi_0 (-L/2, k)^2
),
\eea
since the commutator term on the second line vanishes by the Virial
Theorem. Upon inserting
(\ref{disp0prime}) into the expression (\ref{curr0bis}) for the
edge current, we obtain
\beq
\label{curr0ter}
\langle \psi , V_y  \psi \rangle = \frac{\mathcal{V}_0}{2 B}
\int_{\omega_0^{-1} (\Delta_0)} | \beta_0 (k)|^2  ( \varphi_0 (L/2;k)^2 -
\varphi_0 (-L/2, k)^2 ) ~dk.
\eeq
Consequently, we
are left with the task of estimating the trace of the eigenfunction along
the two boundary components at $x=\pm L/2$.

The key point that allows us to distinguish
these two traces is the following.
The dispersion curves are symmetric about $k=0$ if $V_0(x)$ is an even
function.
Consequently, if a wave function $\psi$ satisfies $\psi = E_0 ( \Delta_0 )
\psi$, we have to study the decomposition of $\psi$
in $k$-space according to the
decomposition $\omega_0^{-1} ( \Delta_0 ) = \omega_0^{-1}( \Delta_0 )_-
\cup
\omega_0^{-1}( \Delta_0 )_+$, where
$\omega_0^{-1} ( \Delta_0 )_{\pm} \equiv \omega_0^{-1} ( \Delta_0 ) \cap
\R_{\pm}$.
These two components
correspond to currents propagating in opposite directions
along the left and right edges of the band, respectively.
To construct a left-edge current, we construct states $\psi$ so that
the coefficients $\beta_0(k)$ in (\ref{coef00})
satisfy
$\mbox{supp} \; \beta_0(k) \subset \omega_0^{-1}(\Delta_0)_- $.
Such a state is spatially concentrated near the left edge $x = -L/2$.
Hence, the contribution to the left-edge current coming
from $\varphi_0 (L/2;k)$ will be exponentially small since the domain $x
\approx L/2$ is in the classically forbidden region
for energies $\omega_0 (k)$, for $k \in \omega_0^{-1}(\Delta_0)_-$.
Consequently,
the contribution to the integral in
(\ref{curr0ter}) will
be exponentially small.
Thus, we prove that if $\psi = E_0 ( \Delta_0) \psi$
is spectrally concentrated in the set $\omega_0^{-1}( \Delta_0)_-$, then the
matrix element
$\langle \psi, V_y \psi \rangle$ is bounded from below by a
constant times $\sqrt{B} \| \psi \|^2$.
Much of our technical work, therefore, is devoted to obtaining lower
bounds on quantities of the form $\mathcal{V}_0 \varphi_0 (\pm L/2;k)^2$
for such left-edge current states.
We also mention that similar results hold for the right-edge current. Of
course, in the
unperturbed case with a symmetric confining potential,
we expect that the net current
across any line $y = C$ is zero for the unperturbed problem.
We will prove this in Proposition 2.1 below.

\subsection{Contents}
This paper is organized as follows. Section 2 is devoted to the estimation
of edge currents for the case of
the sharp confining potential (\ref{potSharp}),
the power function confining potential (\ref{potp}),
and the parabolic confining potential (\ref{potQuad}). In section 3,
the spectral properties of these models are investigated.
Using the Mourre commutator method,
we exhibit a class of potentials $V_1$ (periodic or
decreasing in the $y$-direction) preserving nonempty
absolutely continuous spectrum in intervals lying
between two consecutive Landau levels for the perturbed Hamiltonian
$H_0+V_1$. In section 4,
we address cylinder geometries models and
prove the existence of edge currents for Hamiltonians with
pure point spectrum in this framework.
Appendix 1 in section 5 presents basic properties of the
dispersion curves needed in the proofs. In Appendix 2, section 6,
we collect technical results needed in section 2
for the estimation of edge currents for the power function confining
potential.

\subsection{Acknowledgments}

We thank J.-M. Combes for many discussions on edge currents and their
role in the IQHE. We thank E.\ Mourre for discussions on the commutator
method used in section 3.
We also thank F.\ Germinet, G.-M.\ Graf, and
H.\ Schulz-Baldes for fruitful
discussions.
Some of this work was done when ES was visiting the Mathematics Department
at the University of Kentucky and he thanks the Department for its support.


\renewcommand{\thechapter}{\arabic{chapter}}
\renewcommand{\thesection}{\thechapter}

\setcounter{chapter}{2} \setcounter{equation}{0}

\section{Edge Currents for Two-Edge Geometries}

Many quantum devices can be modeled by a confining potential forcing the
electrons into a strip of infinite extent in one direction. The dynamics of
electrons in an infinite-strip are different from the half-plane cases
treated in \cite{[HS1]}. We study an electron in a strip of width $L >
0$ in the $x$-direction, and unbounded in the $y$-direction.
We consider confining potential
$V_0 (x)$ that are either step functions, or power functions.
After some basic analysis of these
models that is independent of the precise form of the confining potential,
we study edge currents for parabolic confining potential, sharp confining
potential and power function confining potential.

\subsection{Basic Analysis of Two-Edge Geometries}
\label{sub-Basic}
As in \cite{[HS1]}, we study the existence of edge currents for a
general confining potential $V_0(x)$.
We obtain lower bounds on the appropriately localized
velocity along the $y$-direction $V_y=p_y-Bx$.
The strip geometry is a two-edge geometry. Thus, we expect that there is a
current associated with each edge. Classically, these currents propagate
along the edges in opposite directions. For the unperturbed system, one
expects that the net current flow across the line $y=C$, for any $C \in
\R$, to be zero, and we prove this in Proposition \ref{prop-netcurrent}.
Once a perturbation $V_1$ is added, this may no longer be
true, and the persistence of edge currents may depend upon a relationship
between $B$ and $L$.

We continue to use the same notation as in \cite{[HS1]}. That is, we
write $H_0 = H_L(B) + V_0$ for the unperturbed operator. Since we have
translational invariance in the $y$-direction, this operator admits a
direct sum decomposition
\beq
\label{dirsum1}
H_0 = \int^{\oplus}_{\R} ~dk h_0(k).
\eeq
We write $h_0(k)$ for the fibered
operator acting on $L^2 ( \R)$,
where
\beq
\label{fibered1}
h_0(k) = p_x^2 + (k-Bx)^2 + V_0(x),
\eeq
with an even, two-edge confining potential $V_0$. Although
some of our arguments hold for a general confining potential that is
monotone
on the left and the right, we will explicitly treat two cases, the sharp
confining
potential given in (\ref{potSharp}), and the power function confining
potential
given in (\ref{potp}).
We first prove that the total edge current
carried by certain symmetric states of finite energy vanishes.
For this, it is essential that the confining potential
be an {\it even} function.
We consider states of finite energy $\psi$, with $\psi \in E_0( \Delta_n )
\Lp{\R^2}$, for an interval $\Delta_n \subset ( E_n (B) , E_{n+1}(B) )$,
for any $n \geq 0$.
The partial Fourier transform $\hat{\psi}$ of $\psi$
in the $y$-variable
can be expressed in terms of the eigenfunctions $\varphi_j (x;k)$ as
\beq
\label{fopsi1}
\hat{\psi}(x,k)= \Sum_{j=0}^{n} \chi_{\omega_j^{-1}( \Delta_n)} (k) \beta_j
(k) \varphi_j (x;k),
\eeq
or equivalently as
\beq
\label{decomp1}
\psi ( x, y) = \frac{1}{\sqrt{2 \pi}} \Sum_{j=0}^n
\int_\R  e^{iky} ~\chi_{ \omega_j^{-1}(
\Delta_n )} (k) \beta_j (k) \varphi_j (x;k) ~dk ,
\eeq
where the coefficients $\beta_j (k)$ are defined by
\beq
\label{beta1}
\beta_j (k) \equiv \langle \hat{\psi} (\cdot; k ),
\varphi_j ( \cdot ; k) \rangle.
\eeq
and the normalization condition
\beq
\label{stnorm1}
\| \psi\|_{L^2(\R^2)}^2 = \Sum_{j=0}^n  \int_{\omega_j^{-1} (\Delta_n)} |
\beta_j (k) |^2 ~dk.
\eeq
We recall that the
properties of the dispersion curves $\omega_j (k)$ result in the disjoint
decomposition
$\omega_j^{-1} (\Delta_n) = \omega_j^{-1} (\Delta_n)_- \cup \omega_j^{-1}
(\Delta_n)_+$ with $\omega_j^{-1}(\Delta_n)_{\pm} \equiv
\omega_j^{-1}(\Delta_n) \cap \R_{\pm}$.
It is clear from the fact the potential in $h_0(k)$
is centered at $x_0 = k/B$ that the wave function $\psi$ may be more
localized near one edge or
another depending upon the properties of the weights $\beta_j (k)$.
For example, if the $\beta_j (k)$ are supported only by negative wave
numbers $k$, then the wave function will be localized near the left edge.
Such a wave function should carry a net left-edge current. We will prove
this below. We will first prove that if a wave function is symmetrically
localized with respect to the left and right edges, then it carries no net
edge current: The left-edge current cancels the right-edge current.

Let us make the assumption on the confining potential $V_0(x)$ more precise.
In the sequel we assume $V_0$ is an even function which satisfies
simultaneously the two following
conditions:
\beq
\label{condipot}
\left\{ \begin{array}{cl}
(a) & 0 \leq V_0(x) \leq C,\ \forall x \in \R\\
(b) & \lim_{|x| \rightarrow \infty} V_0(x) = C,
\end{array} \right.
\eeq
for some generalized constant $0 < C \leq \infty$.

It is clear that the potential $V_0$ is unbounded at infinity in the case
where $C= \infty$, while it is uniformly bounded by $C$ otherwise.
Actually, each of the particular confining potentials we will consider below
satisfy (\ref{condipot}). Indeed, this is the case for the sharp confining
potential (\ref{potSharp}) for $C=\mathcal{V}_0$, as well as for the power
function confining potential (\ref{potp}) and
the parabolic confining potential (\ref{potQuad}) by taking $C=\infty$.

\vspace{.1in}
\noindent
\begin{proposition}
\label{prop-netcurrent}
Let $V_0(x)$ be a even confining potential
satisfying (\ref{condipot}).
Let $\omega_j (k)$, for $j = 0,1,2, \ldots$, be the dispersion curves for
$h_0 (k)$.
Let $\psi \in E_0( \Delta_n ) \Lp{\R^2}$, as in (\ref{fopsi1}), be a finite
energy state.
Then, the current carried by such a state has the following expression:
\beq
\label{curr1}
\langle \psi,  V_y \psi \rangle = \frac{1}{2} \Sum_{j=0}^n
\int_{\omega_j^{-1} (\Delta_n)_-} (| \beta_j (k) |^2-|\beta_j(-k)|^2)
\omega_j'(k) dk.
\eeq
Henceforth, if $\psi$ is a symmetric state, that is,
$\beta_j (k) = \beta_j (-k)$, for
$j=0,1,\cdots,n$, then the current carried by $\psi$ vanishes:
\beq
\label{nocurr1}
\langle \psi , V_y \psi \rangle = 0 .
\eeq
\end{proposition}

\noindent
\begin{Myproof}
The velocity $V_y = p_y - Bx$ has a Fourier transform that
we write as $\hat{V}_y=\hat{V}_y (k)=k-Bx$.
Using the Fourier decomposition (\ref{fopsi1}),
the matrix element of the velocity operator $V_y$ is
\bea
& &
\langle \psi,  V_y \psi \rangle \label{curr2} \\
& = & \Sum_{j,l=0}^n \int_\R \chi_{\omega_j^{-1} (\Delta_n)}(k)
\chi_{\omega_l^{-1} (\Delta_n)} (k) \overline{\beta}_j (k)
  \beta_l(k)  \langle \varphi_j (\cdot;k), \hat{V}_y(k)
\varphi_l (\cdot;k) \rangle ~dk. \nonumber
\eea
As a consequence of the result of Lemma \ref{lm-cross} below, the
cross-terms in (\ref{curr2}) vanish, at least for $(|\Delta_n| / B)$
sufficiently small, giving
\bea
\label{curr3}
\langle \psi,  V_y \psi \rangle & = &  \Sum_{j=0}^{n} \int_\R
  \chi_{\omega_j^{-1}( \Delta_n)} (k)
  |\beta_j (k)|^2 \langle \varphi_j (\cdot;k ), \hat{V}_y(k) \varphi_j
(\cdot;k
) \rangle ~dk \nonumber \\
&=& \Sum_{j=0}^{n} \int_{-\infty}^0 \chi_{\omega_j^{-1}( \Delta_n)} (k)
   \left\{ |\beta_j (k)|^2 \langle \varphi_j (\cdot;k ), \hat{V}_y (k)
\varphi_j (\cdot;k ) \rangle \right.
\nonumber \\
&& + \left. |\beta_j (-k)|^2 \langle \varphi_j (\cdot;-k ), \hat{V}_y (-k)
\varphi_j
(\cdot; -k ) \rangle \right\} dk,
\eea
where we used the fact, proved in Lemma \ref{lm-symetrie} in Appendix 1,
that the dispersion curves are even functions of
$k$, that is, $\omega_j(k) = \omega_j (-k)$. We also note that
the Hamiltonian $h_0(k)$ commutes with the
operation $P$ that implements $(x,k) \rightarrow (-x,-k)$.
The simplicity of the eigenfunctions then implies that
$P \varphi_j = \pm \varphi_j$.
Hence the last term in the r.h.s. of (\ref{curr3}), $\langle \varphi_j
(\cdot;-k ), \hat{V}_y (-k) \varphi_j
(\cdot; -k ) \rangle$ becomes
\begin{eqnarray*}
\int_{\R} \varphi_j(x;-k)^2 (-k-Bx) dx & = & \int_{\R} \varphi_j(-x;-k)^2
(-k+Bx) dx\\
                                         & = & -\int_{\R} \varphi_j(x;k)^2
(k-Bx) dx\\
                                         & = &  -\langle
\varphi_j (\cdot;k ), \hat{V}_y (k) \varphi_j (\cdot;k ) \rangle,
\end{eqnarray*}
and the result follows from this, (\ref{curr3}) and the Feynman-Hellmann
formula,
\beq
\label{FH}
\omega_j'(k) = 2 \langle \varphi_j (\cdot;k), (k-Bx) \varphi_j (\cdot;k)
\rangle.
\eeq
\end{Myproof}

One of the key points for the proof of Proposition \ref{prop-netcurrent}
and for the estimation of the edge current given below, is the following
Lemma. Its proof relies on the fact (proved in Lemma \ref{lm-separated} in
Appendix 1) the dispersion curves $\omega_j(k)$, $j \in \N$, are separated,
in the sense that
\[ \inf_{k \in \R} | \omega_{l}(k) - \omega_j(k) | >0,\ l \neq j,\
\mbox{for}\ 0 < C < \infty, \]
the same estimate being true if $C=+\infty$ by taking the infimum on any
bounded set instead of $\R$ .

\begin{lemma}
\label{lm-cross}
Let the confining potential $V_0(x)$ be as in Proposition
\ref{prop-netcurrent}.
Let
\beq
\label{defDelta}
\Delta_n
\equiv  [ (2n + a)B , (2n + c)B],\ \mbox{for}\ 1 < a < c
< 3.
\eeq
Then, for any
$j,l = 0,1, \ldots, n$, we have
\beq
\label{nonoverlap2}
\omega_j^{-1} ( \Delta_n ) \cap \omega_l^{-1} ( \Delta_n ) = \emptyset, ~~j
\neq l,
\eeq
provided $c-a$ is sufficiently small.
\end{lemma}
\begin{Myproof}
Let us first consider the case $0 < C < \infty$. In light of Lemma
\ref{lm-separated}, we know that
\[ d_n \equiv \min_{0 \leq j \leq n-1} \inf_{k \in \R} \left(
\omega_{j+1}(k) - \omega_{j}(k) \right) > 0. \]
But any $k \in \omega_j^{-1} ( \Delta_n ) \cap \omega_l^{-1} ( \Delta_n )$
satisfying $0 \leq \omega_{j+1}(k) - \omega_{j}(k) < (c-a) B$, we see this
inequality leads to a contradiction
if $c-a < d_n \slash B$. As a consequence we have $\omega_j^{-1} ( \Delta_n
) \cap \omega_l^{-1} ( \Delta_n ) = \emptyset$ for any $j \neq l$ provided
$| \Delta_n | \slash B < d_n$.

In the case where $C=+\infty$, we deduce from the evenness and the
asymptotic behavior of $\omega_1$ (see Lemmas \ref{lm-symetrie} and
\ref{lm-asyVP} in Appendix 1) there is a real number $k_0>0$ such that
for all $|k| > k_0$ we have
\[ \omega_1(k) > (2n+c) B. \]
The result follows from this by arguing as before with
\[ d_n \equiv \min_{0 \leq j \leq n-1} \inf_{|k| \leq k_0} \left(
\omega_{j+1}(k) - \omega_{j}(k) \right) > 0. \]
\end{Myproof}


\subsection{Edge Currents for a Parabolic Confining Potential}

As a warm up, we address now the model studied by Exner, Joye and Kovarik in
\cite{[EJK]}, where the confining potential is
given by (\ref{potQuad}).
For this model, the electron is confined to a parabolic channel of infinite
extent in the $y$-direction.
For any $E > 0$, the plane $\R^2$ is divided into a classically allowed
region given by $|x| < \sqrt{ E/g } $, and the complementary classically
forbidden region.

Let us define a modified field strength by $B_g \equiv \sqrt{ B^2 + g^2 }$.
The reduced, unperturbed Hamiltonian for the parabolic channel problem is
given by
\bea
\label{unpertham1}
h_0(k) &=& p_x^2 + (k-Bx)^2 + g^2 x^2 \nonumber \\
  &=& p_x^2 + \left( B_g x - \frac{B}{ B_g} k \right)^2 +
    \left( \frac{g}{B_g} \right)^2 k^2 .
\eea
Since this is simply a shifted harmonic oscillator Hamiltonian, it is
completely solvable. The dispersion curves have the following explicit
expression
\beq
\label{Quad-1}
\omega_j (k) = (2j +1) B_g + \left( \frac{g}{B_g} \right)^2 k^2,
\eeq
and the associated normalized eigenfunctions are given by
\beq
\label{Quad-2}
\varphi_j(x;k) = \frac{1}{\sqrt{2^j j!}} \; \left( \frac{B}{\pi}
\right)^{1/4} \; e^{- B_g \slash 2 ( x- (B/B_g^2) k)^2} H_j (\sqrt{B_g} ( x-
(B / B_g^2 ) k)),
\eeq
where $H_j$ is the $j^{\rm \tiny th}$ Hermite polynomial.
The dispersion curves $\omega_j(k)$ being parabolas
with equation (\ref{Quad-1}), the set $\omega_j^{-1}(\Delta_n)$ for the
interval
\beq
\label{defDeltaG}
\Delta_n \equiv [(2n+a)B_g,(2n+c)B_g],\ 1<a<c<3,
\eeq
is explicitly known:
\beq
\label{Quad-3}
\omega_j^{-1}(\Delta_n)_-=[-k_j^{(n)}(c),-k_j^{(n)}(a)],
\eeq
with
\begin{equation}
\label{Quad-4}
k_j^{(n)}(x) \equiv \frac{B_g^{\troisdemip}}{g} \sqrt{2(n-j)+x - 1},\ x=a,
c.
\end{equation}
Henceforth,
\[ -\omega_j'(k) = - 2 \left( \frac{g}{B_g} \right)^2 k \geq 2 \left(
\frac{g}{B_g} \right)^2 k_j^{(n)}(a), \]
for each $k \in \omega_j^{-1}(\Delta_n)_-$, which leads to
\beq
\label{Quad-5}
-\omega_j'(k) \geq 2 \left( \frac{g}{\sqrt{B_g}} \right) \sqrt{2(n-j)+a-1},\
k \in \omega_j^{-1}(\Delta_n)_-,
\eeq
according to (\ref{Quad-4}).

Let us consider a state of finite energy $\psi$, with $\psi \in E_0(
\Delta_n )
\Lp{\R^2}$, whose Fourier coefficients $\beta_j(k)$, $j=0,1,\ldots,n$, are
defined as in (\ref{beta1}).
We assume in addition there is a constant $\gamma >0$ such that the
$\beta_j(k)$ satisfy the following condition:
\beq
\label{nonsympsi}
| \beta_j(k) |^2 \geq (1+\gamma^2) | \beta_j(-k) |^2,\ k \in
\omega_j^{-1}(\Delta_n)_-, j=0,1,\cdots,n.
\eeq
Thus $| \beta_j(k) |^2 - | \beta_j(-k) |^2 \geq \gamma^2 \slash (1+\gamma^2)
| \beta_j(k) |^2$
for all $j=0,1,\ldots,n$ and $k \in \omega_j^{-1}(\Delta_n)_-$, so we get
\beq
\label{nonsympsi2}
\sum_{j=0}^n \int_{\omega_j^{-1}(\Delta_n)_-} | \beta_j(k) |^2 dk \geq
\frac{1 + \gamma^2}{2 + \gamma^2} \| \psi \|^2,
\eeq
from the normalization condition (\ref{stnorm1}). It follows readily from
this and from the expression (\ref{curr1}) of the total current carried by
the state $\psi$,
\[
\langle \psi,  V_y \psi \rangle = \frac{1}{2} \Sum_{j=0}^n
\int_{\omega_j^{-1} (\Delta_n)_-} (| \beta_j (k) |^2-|\beta_j(-k)|^2)
\omega_j'(k) dk,
\]
together with the estimate (\ref{Quad-5}), that
\beq
\label{Quad-6}
-\langle \psi, V_y \psi \rangle \geq \frac{\gamma^2}{2+\gamma^2}
\sqrt{(a-1)} \frac{g}{B_g^{\demip}}.
\eeq
Notice that the lower bound to the current in (\ref{Quad-6}) is actually of
size $B^{\demip}$ since $g$ has the same dimension as $B$.


\subsection{Estimation of the Edge Current for a Strip}
\label{sub-EstEdgeCur}
We turn now to the
estimation of the left-edge current for a strip of width $L>0$. Namely, we
assume the confining potential $V_0$ is an even function satisfying
(\ref{condipot}) and such that
\beq
\label{StripConfPot}
V_0(x) \chi_{ \{ |x| < L \slash 2 \} }(x) = 0.
\eeq
We want to estimate the total current along both edges, carried by
appropriately chosen states $\psi$. That is, we want to obtain a lower
bound on the matrix element of the localized velocity operator
(\ref{curr1}), carried by a state $\psi \in E_0(\Delta_n) \Lp{\R^2}$
associated to the energy interval $\Delta_n
\subset ( E_n(B), E_{n+1} (B) )$.
Much of the technical work in this paper is devoted to bounding
$(-\omega_j'(k))$ from below, uniformly for $k$ in $\omega_j^{-1}
(\Delta_n)_-$.
\begin{lemma}
\label{lm-main}
Let $\Delta_n$ be as in Lemma \ref{lm-cross} and $\beta_j$,
$j=0,1,\cdots,n$, be defined by (\ref{beta1}). Then, there is a constant
$C_n > 0$ independent of $B$ such that
\beq
\label{derdiscur1}
-\omega_j'(k) \geq C_n (a-1)^2 (3-c)^2 B^{\demip},\ k \in
\omega_j^{-1}(\Delta_n)_-,
\eeq
provided $B$ is large enough and $V_0$ satisfies one the two following
\mbox{conditions:}
\begin{itemize}
\item $V_0$ is the sharp confining potential defined by (\ref{potSharp}) and
$\mathcal{V}_0 \geq 2(2n+c)B$,
\item $V_0$ is a power function as in (\ref{potp}) and $\mathcal{V}_0 \geq
(2n+c)B^{(p+2) \slash 2}$.
\end{itemize}
Moreover $C_n$ does not depend on $\mathcal{V}_0$ in the case of the sharp
confining potential, while
$C_n =\tilde{C}_n \slash v$ where $v =  (2n+c) B^{-(p+2) \slash 2}
\mathcal{V}_0 \geq 1$ and $\tilde{C}_n$ is independent of $\mathcal{V}_0$
for the power function confining potential.
\end{lemma}
\begin{Myproof}
Inserting the commutator formula
\beq
\label{comm1}
\hat{V}_y(k) = (k-Bx) \equiv \frac{-i}{2B} [ p_x , h_0 (k) ] +
\frac{1}{2B} V_0'
\eeq
for $\hat{V}_y (k)$ into (\ref{FH}), we obtain
two terms. Due to the Virial Theorem, the term involving the commutator
$[p_x, h_0(k)]$ vanishes as in the one-edge case, giving:
\beq
\label{FH2}
\omega_j'(k) =  \frac{1}{2B} \langle \varphi_j (\cdot;k), V_0' \varphi_j
(\cdot;k) \rangle.
\eeq
The end of the proof also consists in bounding the remaining term
$\langle \varphi_j (\cdot;k), V_0' \varphi_j (\cdot;k) \rangle$ from above
by a (negative) constant times $B^{3 \slash 2}$.
This technical computation is postponed to section \ref{sub-Sharp-Estimate}
for the sharp confining potential and to
section \ref{sub-p-Estimate} for the power function confining potential. In
both cases the technique used is based on Lemmas \ref{lm-WNE}
and \ref{lm-SED} given in section \ref{sub-WNE} below.
\end{Myproof}

In light of (\ref{curr1}) and Lemma \ref{lm-main}, let us see now the
current carried by a state $\psi$, whose coefficients $\beta_j(k)$,
$j=0,1,\cdots,n$, are mostly supported on the set of negative wave numbers
$k$, is of size $B^{\demip}$.

\begin{theorem}
\label{thm-cur}
Let $\Delta_n$ and $V_0$ be as in Lemma \ref{lm-main}, and $\psi$ satisfy
the condition (\ref{nonsympsi}): There is $\gamma>0$ such that
\[ | \beta_j(k) |^2 \geq (1+\gamma^2) | \beta_j(-k) |^2,\ k \in
\omega_j^{-1}(\Delta_n)_-, j=0,1,\cdots,n. \]
Then for sufficiently large $B$, we have
\beq
\label{curr4}
-\langle \psi, V_y \psi \rangle \geq \frac{\gamma^2}{2+\gamma^2} C_n (a-1)^2
(3-c)^2 B^{\demip} \| \psi \|^2,
\eeq
where $C_n$ is the constant defined in Lemma \ref{lm-main}.
\end{theorem}
\begin{Myproof}
By recalling the estimate (\ref{nonsympsi2}),
\[ \sum_{j=0}^n \int_{\omega_j^{-1}(\Delta_n)_-} | \beta_j(k) |^2 dk \geq
\frac{1 + \gamma^2}{2 + \gamma^2} \| \psi \|^2, \]
which derives from (\ref{nonsympsi}) together with the normalization
condition (\ref{stnorm1}), the result immediately follows from (\ref{curr1})
and (\ref{derdiscur1}).
\end{Myproof}


\subsection{Bounding the Right Current Term}
\label{sub-WNE}

As in section \ref{sub-EstEdgeCur} we assume the confining potential $V_0$
is an even function satisfying (\ref{condipot}) and (\ref{StripConfPot}).
This is the case for the step function confining potential (\ref{potSharp})
and the power function confining potential (\ref{potp}) we will consider
below.

For any $j=0,1,\ldots,n$ it is clear from the definition of
$\omega_j^{-1}(\Delta_n)_-$ that $\sup \omega_j^{-1}(\Delta_n)_- \leq 0$.
Actually, we establish in Lemma \ref{lm-WNE}
this supremum is bounded by a number arbitrarily close to $(-BL) \slash 2$,
provided the magnetic strength $B$ is taken sufficiently large.
Consequently, the region $x \geq 0$ is in the classically forbidden zone for
energies $\omega_j(k)$, $k \in \omega_j^{-1}(\Delta_n)_-$, at least in the
intense magnetic field regime.
This is because the parabolic part of the effective potential
\beq
\label{EffPot}
W_j(x;k) \equiv (k-Bx)^2 + V_0(x) - \omega_j(k),
\eeq
is centered at the coordinate $k \slash B$.

Henceforth the eigenfunctions $\varphi_j(.;k)$ of $h_0(k)$ are exponentially
decaying in the region $x \geq 0$ for all $k \in \omega_j^{-1}(\Delta_n)_-$.
This is not true in the region $x \leq 0$. Due to the evenness of $V_0$,
$\int_{\R_+} V_0'(x) \varphi_j(x;k)^2 dx$ is also expected to be small
relative to $\int_{\R_-} V_0'(x) \varphi_j(x;k)^2 dx$, so
\[ \omega_j'(k) \approx \frac{1}{2B} \int_{\R_-} V_0'(x) \varphi_j(x;k)^2
dx, \]
according (\ref{FH2}).
This remark is made precise below. Namely we state in Lemma \ref{lm-SED}
that the remaining term $\int_{\R_-} V_0'(x) \varphi_j(x;k)^2 dx$ is bounded
by a constant times $B$ and we establish
in sections \ref{sub-Sharp-Estimate}-\ref{sub-p-Estimate} for the step
function confining potential (\ref{potSharp}) and
the power function confining potential (\ref{potp}), the main term
$\int_{\R_-} V_0'(x) \varphi_j(x;k)^2 dx$ is of size
$B^{3 \slash 2}$.

\subsubsection{Wave Numbers Estimate}

\begin{lemma}
\label{lm-WNE}
Let $\Delta_n$ be as in Lemma \ref{lm-cross} and $V_0$ be an even function
satisfying (\ref{condipot}) and (\ref{StripConfPot}). Then, any given
$\alpha >2$, there is $B_{\alpha} \geq 1$ such that
\[ \omega_j^{-1}(\Delta_n)_- \subset (-\infty,-BL \slash \alpha), \]
for all $B \geq B_{\alpha}$ and $\mathcal{V}_0 > 0$.
\end{lemma}
\begin{Myproof}
Let $\theta_{\epsilon}$ be a real valued, even and twice continuously
differentiable function in $\R$, such that
\[ \theta_{\epsilon}(x) = \left\{ \begin{array}{cl}
    1 & \mbox{if}\ -L \slash 2 + \epsilon \slash 2 \leq  x \leq 0\\
    0 & \mbox{if}\ x < -L \slash 2 + \epsilon \slash 4, \end{array} \right.
\]
for some $\epsilon$ in  $(0,L \slash 2)$.
The function $\theta_{\epsilon}(x) \psi_n(x;k)$ (where $\psi_n(x;k)$ still
denotes the $n^{\tiny \mbox{th}}$ normalized eigenfunction of
$h_L(k)=p_x^2+(k-Bx)^2$)
obviously belongs to the domain of $h_0(k)$. Moreover, the supports of $V_0$
and $\theta_{\epsilon}$ being disjoint, the following identity holds true:
\begin{eqnarray*}
( h_0(k) - (2n+1)B ) \theta_{\epsilon}(x) \psi_n(x;k) & = & [ h_0(k),
\theta_{\epsilon} ] \psi_n(x;k)\\
& = & -(\theta_{\epsilon}''+2i \theta_{\epsilon}' p_x) \psi_n(x;k).
\end{eqnarray*}
This immediately entails:
\begin{equation}
\label{Di1}
\| ( h_0(k) - (2n+1)B ) \theta_{\epsilon} \psi_n(.;k) \| \leq \|
\theta_{\epsilon}'' \psi_n(.;k) \| + 2 \| \theta_{\epsilon}' p_x \psi_n(x;k)
\|.
\end{equation}
Let us suppose now that $k \slash B \in [-L \slash 2 + \epsilon, 0]$.
Then, using the explicit expression (\ref{psim}) of $\psi_n(.;k)$ and
bearing in mind the vanishing of
$\theta_{\epsilon}'$ outside $[-L \slash 2 + \epsilon \slash 4,-L \slash 2 +
\epsilon \slash 2] \cup [L \slash 2 - \epsilon \slash 2, L \slash 2 -
\epsilon \slash 4]$,
it is possible to find two constants $\alpha_n$ and $\beta_n$ independent of
$B$, $\mathcal{V}_0$ and $\epsilon$, such that:
\[ \left\{
\begin{array}{l}
\| \theta_{\epsilon}'' \psi_n(.;k) \| \leq  \alpha_n  \epsilon^{-3 \slash 2}
B^{1 \slash 4} \expo{- B \slash 8 \epsilon^2}\\
\| \theta_{\epsilon}' p_x \psi_n(x;k) \| \leq \beta_n  \epsilon^{-1 \slash
2} B^{1 \slash 4} (B^{\demip} + \epsilon B) \expo{- B \slash 8 \epsilon^2}.
\end{array} \right. \]
This, combined with (\ref{Di1}), involves
\beq
\label{Di1bis}
\| ( h_0(k) - (2n+1)B ) \theta_{\epsilon} \psi_n(.;k) \| \leq \gamma_n B^{1
\slash 4} \epsilon^{-3 \slash 2} (1+ \epsilon B^{1 \slash 2} + \epsilon^{2}
B ) \expo{- B \slash 8 \epsilon^2},
\eeq
where we have set $\gamma_n=\alpha_n + 2 \beta_n$.\\
Next, by performing the change of variable $y=B^{\demip}(x-k \slash B)$ in
the
integral $\int_{\R} \theta_{\epsilon}^2 \psi_n(x;k)^2 dx$ we get
\begin{eqnarray*}
2^n n! \sqrt{\pi} \| \theta_{\epsilon} \psi_n(.;k) \|^2 & \geq &
\int_{B^{\demip}(-L \slash 2 + \epsilon \slash 2 - k \slash
B)}^{B^{\demip}(L \slash 2 - \epsilon \slash 2 - k \slash B)}
H_n(y)^2 \expo{-y^2} dy \\
& \geq & \int_{0}^{L \slash  4} H_n(y)^2 \expo{-y^2} dy > 0,
\end{eqnarray*}
for all $B \geq 1$.
In light of (\ref{Di1bis}), we see there is also a constant $C_n$
independent of $B$, $\mathcal{V}_0$ and $\epsilon$ such that
\[ \dist( \sigma(h_0(k)), (2n+1)B ) \leq C_n B^{1 \slash 4} \epsilon^{-3
\slash 2} (1+ \epsilon B^{1 \slash 2} + \epsilon^{2} B ) \expo{- B \slash 8
\epsilon^2}, \]
provided $B$ is sufficiently large.
This, combined with the simplicity of the $\omega_m(k)$, entails
\[ \omega_n(k) \leq (2n+1) B + C_n B^{1 \slash 4} \epsilon^{-3 \slash 2}
(1+\epsilon B^{1 \slash 2} + \epsilon^{2} B ) \expo{- B \slash 8
\epsilon^2},\]
proving that $\omega_n(k)$ can be made smaller than $(2n+a)B$ by taking $B$
sufficiently large. Hence we have shown that
\beq
\label{Di2}
\omega_j(k) \notin \Delta_n,\ j=0,1,\ldots,n,\ k \slash B \in [-L \slash 2 +
\epsilon,0],
\eeq
and the result follows from (\ref{Di2}) for all $k \in [-BL \slash \alpha,
0]$ by taking $\epsilon = (\alpha -2) \slash (2 \alpha) L$.
\end{Myproof}

\subsubsection{Trace Function Estimate in the Classically Forbidden Zone}
The main consequence of the preceding Lemma is the positivity of the
effective potential $W_j(x;k)$ defined by (\ref{EffPot}) for $k \in
\omega_j^{-1}(\Delta_n)_-$
in the region $x \geq 0$.
Indeed, we know from Lemma \ref{lm-WNE} we can make $B$ large enough so
$\omega_j^{-1}(\Delta_n)_- \subset (-\infty,-BL \slash 3)$, and consequently
$W_j(x;k) \geq B^2 L^2 \slash 36 - (2n+c) B$ for all $x \geq -L \slash 6$
and $k \in \omega_j^{-1}(\Delta_n)_-$. Whence there is necessarily $B_0>0$
such that:
\beq
\label{tracebound0}
W_j(x;k) \geq \left( \frac{B L}{8} \right)^2 > 0,\ x \geq - L \slash 6,\ k
\in \omega_j^{-1}(\Delta_n)_-,\  B \geq B_0.
\eeq
The eigenfunction $\varphi_j(.;k)$ being an $\Hp{\R}$-solution to the
differential equation $\varphi''(x)=W_j(x;k) \varphi(x)$, is also
exponentially decaying in the region $x \geq - L \slash 6$. Namely, we have
\beq
\label{tracebound1}
0 \leq \varphi_j(t;k) \leq \varphi_j(s;k) \expo{-\int_s^t \sqrt{W_j(x;k)}
dx},\ -L \slash 6 \leq s \leq t,
\eeq
from Proposition 8.2 in \cite{[HS1]}. This estimate is the main tool to
bound $\int_{\R_+} V_0'(x) \varphi_j(x;k)^2 dx$ as in Lemma \ref{lm-SED}.
The proof consists in relating
this integral to $\int_{\R_+} (Bx-k) \varphi_j(x;k)^2 dx$ through the
generalized expression (\ref{FH1}) of the Feynman-Hellmann relation.
Concerning, $\int_{\R_+} (Bx-k) \varphi_j(x;k)^2 dx$, upon choosing $B$ is
sufficiently large, we actually have:
\beq
\label{preSED}
\int_{0}^{+\infty} (Bx-k) \varphi_j(x;k)^2 dx \leq \frac{BL}{2} \expo{-BL^2
\slash 24},\ k \in \omega_j^{-1}(\Delta_n)_-.
\eeq
The proof of (\ref{preSED}) is based on the estimate (\ref{tracebound1}) and
consists in 2 steps.\\
{\it First Step.} In light of Lemma \ref{lm-WNE}, we choose $B$ large enough
so
\[ 0 \leq Bt-k  \leq 2 W_j^{\demip}(t;k),\ t \geq 0,\ k \in
\omega_j^{-1}(\Delta_n)_-,\]
then we combine this estimate with (\ref{tracebound1}) written with $s=0$
and integrate the obtained inequality over $(0,+\infty)$, getting:
\beq
\label{tracebound1bis}
\int_{0}^{+ \infty} (Bt-k) \varphi_j(t;k)^2 dt \leq  \varphi_j(0;k)^2.
\eeq
{\it Second Step.} We insert (\ref{tracebound0}) in (\ref{tracebound1})
written with $t=0$, square,
\[ \varphi_j(0;k)^2 \expo{-(B L \slash 4) s} \leq \varphi_j(s;k)^2,\  -L
\slash 6  \leq s \leq 0, \]
then we integrate the obtained inequality with respect to $s$ over the
interval $(-L \slash 6,0)$.
Thus, using the normalization condition $\| \varphi_j(.;k) \| = 1$, we
obtain:
\beq
\label{tracebound2}
\varphi_j(0;k)^2\leq \frac{B L}{2} \expo{-BL^2 \slash 24},\ k \in
\omega_j^{-1}(\Delta_n)_-.
\eeq
Finally (\ref{preSED}) follows from (\ref{tracebound1bis}) and
(\ref{tracebound2}).

Armed with (\ref{preSED}) we turn now to establishing the main result of
this section.

\subsubsection{Bounding the Right Current Term}
\begin{lemma}
\label{lm-SED}
Let $\Delta_n$ and $V_0$ be as in Lemma \ref{lm-WNE}.
Then, there is $B_1>0$ and a constant $\gamma(n,j)>0$ independent of $B$ and
$\mathcal{V}_0$, such that:
\beq
\label{eq-SED}
\int_{\R_+} V_0'(x) \varphi_j(x;k)^2 dx \leq \gamma(n,j) B,\ k \in
\omega_j^{-1}(\Delta_n)_-,\ B \geq B_1.
\eeq
\end{lemma}
\begin{Myproof}
Let $\rho \in C^{3}(\R)$ be a bounded real-valued function and $A$ denote
the selfadjoint operator $\rho(x) p_x + p_x \rho(x)$ in $\Lp{\R}$, with
domain $\Hp{\R}$. Any function $\varphi$ in the domain of $h_0(k)$ belonging
to $\Hp{\R}$,
$\pscal{\Lp{\R}}{[A,h_0(k)] \varphi}{\varphi}$ can be defined as
$\pscal{\Lp{\R}}{h_0(k) \varphi}{A\varphi}-\pscal{\Lp{\R}}{A\varphi}{
h_0(k)\varphi}$, and
standard computations provide:
\begin{eqnarray}
\pscal{\Lp{\R}}{-i [A,h_0(k)] \varphi}{\varphi}
& = & + 4 \pscal{\Lp{\R}}{\rho' \varphi'}{\varphi'} -4B \pscal{\Lp{\R}}{\rho
(Bx-k)\varphi}{\varphi} \nonumber \\
& & - 2 \pscal{\Lp{\R}}{\rho V_0' \varphi}{\varphi} -
\pscal{\Lp{\R}}{\rho''' \varphi}{\varphi}. \label{FH1}
\end{eqnarray}
Here $\pscal{\Lp{\R}}{\rho V_0' \varphi}{\varphi}$ means
$\mathcal{V}_0 (\rho(L \slash 2) \varphi^2(L \slash 2)- \rho(-L \slash
2)\varphi^2(-L \slash 2))$ in the case where $V_0$ is the sharp confining
potential (\ref{potSharp}).
When $\varphi$ is an eigenfunction $\varphi_j(.;k)$ of $h_0(k)$, the scalar
product $\pscal{\Lp{\R}}{-i [A,h_0(k)] \varphi}{\varphi}$ vanishes according
to the Virial Theorem.
Henceforth by taking $\rho$ such that $\rho(x)=0$ if $x \leq 0$, and
$\rho(x)=1$ if $x\geq L \slash 2$, we deduce from (\ref{FH1})
\begin{eqnarray*}
& & 2 \pscal{\Lp{\R}}{\rho V_0' \varphi_j(.;k)}{\varphi_j(.;k)} \\
& \leq & \| \rho''' \|_{\infty} \int_0^{L \slash 2} \varphi_j(x;k)^2 dx + 4B
  \| \rho \|_{\infty} \int_0^{\infty} (Bx-k) \varphi_j(x;k)^2 dx \\
& & + 4 \| \rho' \|_{\infty} \int_0^{L \slash 2} \varphi_j'(x;k)^2 dx.
\end{eqnarray*}
Hence the result follows from this, (\ref{preSED}), together with the basic
inequality $\int_{\R} \varphi_j'(x;k)^2 dx \leq (2n+c)B$.
\end{Myproof}

Notice that (\ref{eq-SED}) actually reduces to
\[ \begin{array}{cll}
      (i) & \mathcal{V}_0 \varphi_j(L \slash 2;k)^2 \leq \gamma(n,j) B &
\mbox{if $V_0$ is given by (\ref{potSharp})}\\
      (ii) & \mathcal{V}_0 \int_{(L \slash 2,+\infty)}  (x-L \slash 2)^{p-1}
\varphi_j(x;k)^2 dx \leq \gamma(n,j) B & \mbox{if $V_0$ is given by
(\ref{potp})}.
   \end{array} \]
These two estimates are useful for the two following sections.


\subsection{The Sharp Confining Potential}
\label{sub-Sharp-Estimate}
The sharp confining potential $V_0$, defined by (\ref{potSharp})
confines particles with energy less than $\mathcal{V}_0$
to the strip $-L/2 \leq x \leq L/2$. For this model, we have
\[ V_0'(x)=\mathcal{V}_0 ( \delta(x+ \demiLp) - \delta(x- \demiLp)) \]
in the distributional sense, so the derivative of the $j^{\tiny \mbox{th}}$
dispersion curve can be expressed as
\beq
\label{derdiscur2}
\omega_j'(k) =  -\frac{ \mathcal{V}_0}{2B} \left( \varphi_j(-\demiLp;k)^2 -
\varphi_j(\demiLp;k)^2 \right),
\eeq
according to (\ref{FH2}).

The case of the left part of the current is treated by Lemma \ref{lm-SED}:
Upon taking $B$ sufficiently large we have,
\beq
\label{SEDSharp}
\mathcal{V}_0 \varphi_j(L \slash 2;k)^2 \leq \gamma(n,j) B,\ k \in
\omega_j^{-1}(\Delta_n)_-,
\eeq
the constant $\gamma(n,j)>0$ being independent of $B$ and $\mathcal{V}_0$.

We turn now to computing a lower bound on the trace term $\mathcal{V}_0
\varphi_j(-\demiLp;k)^2$. This will require several steps.\\

\noindent
{\bf Step 1 : Harmonic Oscillator Eigenfunction Comparison Revisited}\\
The proof of Lemma 2.2 in \cite{[HS1]} (based on the properties of the
eigenfunctions $\psi_m(.;k)$ of the
harmonic oscillator $h_L(k)=p_x^2 + (Bx-k)^2$) applying without change to
the case of the strip geometry examined here,
the following estimate,
\beq
\label{Sharp-lowerbound}
| \pscal{}{\varphi_j(.;k)}{V_0 P_n \varphi_j(.;k)} | \geq \frac{1}{2(n+1)B}
( \omega_j(k)-E_n(B) ) (E_{n+1}(B) - \omega_j(k) ),
\eeq
holds for all $k \in \omega_j^{-1}(\Delta_n)_-$.
We recall that $P_n$ denotes the projection on the eigenspace spanned by the
first $n$ eigenfunctions $\psi_m(.;k)$ of $h_L(k)$,
\beq
\label{Pn}
P_n \varphi_j(x;k) \equiv \sum_{m=0}^n \alpha_m^{(j)}(k) \psi_m(x;k),
\eeq
with
\beq
\label{}
\alpha_m^{(j)}(k) \equiv \pscal{}{\varphi_j(.;k)}{\psi_m(.;k)},
\eeq
and that the explicit expression of $\psi_m(x;k)$ is
\beq
\label{psim}
\psi_m(x;k)=\frac{1}{\sqrt{2^m m!}} \left( \frac{B}{\pi} \right)^{\quartp}
H_m( \sqrt{B} ( x-k \slash B ) ) \expo{-B \slash 2 (x - k \slash B)^2},
\eeq
where $H_m$ denotes the $m^{\tiny \mbox{th}}$ Hermite polynomial function as
in \cite{[HS1]}.

The strategy consists in computing an upper bound on $|
\pscal{}{\varphi_j(.;k)}{V_0 P_n \varphi_j(.;k)} |$, involving the trace
$\mathcal{V}_0^2 \varphi_j(-L \slash 2;k)^2$. To do that, we first calculate
the scalar product $\pscal{}{\varphi_j(.;k)}{V_0 P_n \varphi_j(.;k)}$ by
expanding $P_n \varphi_j(.;k)$ as in (\ref{Pn}):
\beq
\label{estima1}
| \pscal{}{\varphi_j(.;k)}{V_0 P_n \varphi_j(.;k)} | \leq \mathcal{V}_0
\sum_{m=0}^n | \alpha_m^{(j)}(k) | \int_{|x| \geq L \slash 2}
|\varphi_j(x;k)| |\psi_m(x;k) | dx.
\eeq
For this model, the set $| x | > L \slash 2$ is the classically forbidden
region for electrons with energy less than $\mathcal{V}_0$, so
\[ 0 \leq \varphi_j(x;k) \leq \varphi_j(\pm L \slash 2;k) \expo{\mp
\sqrt{\mathcal{V}_0-\omega_j(k)} (x \mp L \slash 2)},\ \pm x \geq L \slash
2, \]
from Proposition 8.3 in \cite{[HS1]}.
Henceforth by substituting the corresponding exponentially decreasing term
for $\varphi_j(.;k)$ in (\ref{estima1}), we obtain
\bea
& & | \pscal{}{\varphi_j(.;k)}{V_0 P_n \varphi_j(.;k)} | \nonumber\\
& \leq & \mathcal{V}_0 \sum_{m=0}^n | \alpha_m^{(j)}(k) | \left(
(I_{m,-}^{(j)}) \varphi_j(-L \slash 2;k) + (I_{m,+}^{(j)}) \varphi_j(L
\slash 2;k) \right), \label{estima2}
\eea
where we have set
\beq
\label{Ipm}
I_{m,\pm}^{(j)} \equiv \int_{\pm x \geq L \slash 2} | \psi_m(x;k) |
\expo{\mp \sqrt{\mathcal{V}_0 - \omega_j(k)} (x \mp L \slash 2)} dx.
\eeq
\noindent
{\bf Step 2 : Trace Function Estimate} \\
In view of bounding the integrals $I_{m,\pm}^{(j)}$ we first define the
constant
\beq
\label{defmaHm}
\mathcal{H}_m \equiv \sup_{u \in \R} H_m(u) \expo{-u^2 \slash 2}.
\eeq
Then we substitute the following estimate
\beq
\label{boundpsim}
| \psi_m(x;k) | \leq \left( \frac{B}{\pi} \right)^{\quartp}
\frac{\mathcal{H}_m}{\sqrt{2^m m!}},
\eeq
which obviously follows from (\ref{psim}) and (\ref{defmaHm}), for $|
\psi_m(x;k) |$ in (\ref{Ipm}), and get:
\beq
\label{Ipm2}
I_{m,\pm}^{(j)} \leq \left( \frac{B}{\pi} \right)^{\quartp}
\frac{\mathcal{H}_m}{\sqrt{2^m m!}} \frac{1}{\sqrt{\mathcal{V}_0 -
\omega_j(k)}}.
\eeq
Now combining (\ref{estima2}) with (\ref{Ipm2}), we obtain
\bea
& & | \pscal{\Lp{\R^2}}{\varphi_j(.,k)}{V_0 P_n \varphi_j(.,k)} |
\label{estima3} \\
& \leq & \frac{\mathcal{V}_0}{\sqrt{\mathcal{V}_0 - \omega_j(k)}}  \left(
\frac{B}{\pi} \right)^{\quartp} \left( \sum_{m=0}^n
\frac{\mathcal{H}_m}{\sqrt{2^m m!}} | \alpha_m^{(j)}(k) | \right)
\left( \varphi_j(-L \slash 2;k) + \varphi_j(L \slash 2;k) \right). \nonumber
\eea
Let us define the constant $\mathcal{H}^{(n)}$ by
\beq
\label{defn1}
\mathcal{H}^{(n)} \equiv \left( \Sum_{m \leq n} \frac{\mathcal{H}_m^2 }{ 2^m
m!} \right)^{1/2}.
\eeq
Then we apply the Cauchy-Schwarz inequality to the sum in (\ref{estima3}),
and
use the normalization condition
\[ \sum_{m=0}^n |\alpha_m^{(j)}(k) |^2 = \| P_n \varphi_j(\cdot;k) \|^2 \leq
1, \]
so we end up getting:
\bea
& & | \pscal{\Lp{\R^2}}{\varphi_j(.,k)}{V_0 P_n \varphi_j(.,k)} |
\label{estima4} \\
&\leq & \frac{\mathcal{V}_0}{\sqrt{\mathcal{V}_0 - \omega_j(k)}} \left(
\frac{B}{\pi} \right)^{\quartp} \mathcal{H}^{(n)}
\left( \varphi_j(-L \slash 2;k) + \varphi_j(L \slash 2;k) \right). \nonumber
\eea
Thus (\ref{estima4}) combined with (\ref{SEDSharp}) and
(\ref{Sharp-lowerbound}) provides
\begin{eqnarray*}
& & \mathcal{V}_0^{\demip} \varphi_j(-L \slash 2;k)  \\
& \geq  & \frac{\pi^{\quartp}}{2(n+1) \mathcal{H}^{(n)}} \left(
1-\frac{\omega_j(k)}{\mathcal{V}_0} \right)^{\demip}  (a-1)(3-c) B^{3 \slash
4} - \sqrt{\alpha(n,j)} B^{1 \slash 4},
\end{eqnarray*}
so there is a constant $D(n,j)$ independent of $B$ and $\mathcal{V}_0$ such
that
\beq
\label{Sharp-trace1}
\mathcal{V}_0^{\demip} \varphi_j(-L \slash 2;k)\geq D(n,j) (a-1) (3-c) B^{3
\slash 4},
\eeq
for all $j=0,1,\cdots,n$ and $k \in \omega_j^{-1}(\Delta_n)_-$. This
estimate holds provided $B$ is sufficiently large, $\mathcal{V}_0 \geq
2(2n+c)B$ and $|\Delta_n|$ is small enough.

Now, the bound (\ref{derdiscur1}) on the derivative $\omega_j'(k)$ follows
from (\ref{derdiscur2}), (\ref{SEDSharp}) and (\ref{Sharp-trace1}).


\subsection{The Power Function Confining Potential}
\label{sub-p-Estimate}
The second case we consider is the one for which the
confining potential is a power function of $x$ alone and is
given by (\ref{potp}).
Due to (\ref{FH2}), the derivative of the $j^{\tiny \mbox{th}}$ dispersion
curve, $j=0,1,\ldots,n$, in the particular case of the power function
confining potential (\ref{potp}) has the following expression
\beq
\label{pderivative}
\omega_j'(k) = -p \frac{\mathcal{V}_0}{2B} \left( I_p^{(j)}(k) -
I_p^{(j)}(-k) \right),
\eeq
where
\beq
\label{defIpj}
I_p^{(j)}(k) = \int_{-\infty}^{-L \slash 2} (-x-L \slash 2)^{p-1}
\varphi_j(x;k)^2 dx.
\eeq
Here we used the symmetry property $\varphi_j(-x;k)^2 = \varphi_j(x;-k)^2$
of the eigenfunctions established in Lemma \ref{lm-symetrie}.

The second term in (\ref{pderivative}) is treated by Lemma \ref{lm-SED}:
There is a constant $\gamma(n,j)>0$ independent of $B$ and $\mathcal{V}_0$
such that we have
\beq
\label{Ipjnegligible}
0 \leq \mathcal{V}_0 I_p^{(j)}(-k) \leq \gamma(n,j) B,\ k \in
\omega_j^{-1}(\Delta_n)_-.
\eeq
Actually (\ref{Ipjnegligible}) holds true provided $B$ is taken sufficiently
large.

We turn now to estimating from below the integral $I_p^{(j)}(k)$ defined in
(\ref{defIpj}).
We follow the calculation of section \ref{sub-Sharp-Estimate}.\\

\noindent
{\bf Step 1 : Harmonic Oscillator Eigenfunction Comparison}\\
As in section \ref{sub-Sharp-Estimate} the starting point of the method is
the estimate (\ref{Sharp-lowerbound}). Namely, for any $k \in
\omega_j^{-1}(\Delta_n)$, we have
\[ | \langle \varphi_j ( \cdot;k), V_0 P_n \varphi_j (\cdot;k) \rangle|
\geq
\frac{1}{2B(n+1)} ( \omega_j(k) - E_n(B) )( E_{n+1}(B) - \omega_j(k)), \]
where $P_n$ still denotes the projection on the eigenspace spanned by the
first $n$ eigenfunctions $\psi_m(.;k)$ of the harmonic oscillator
Hamiltonian $h_L (k)=p_x^2+(Bx-k)^2$.
The strategy consists in computing an upper bound on
$| \langle \varphi_j ( \cdot;k), V_0 P_n \varphi_j (\cdot;k) \rangle|$,
involving $\mathcal{V}_0 I_p^{(j)}(k)$.
To do that, we expand $P_n  \varphi_j(.;k)$ as in (\ref{Pn}) in $\langle
\varphi_j ( \cdot;k), V_0 P_n \varphi_j (\cdot;k) \rangle$, getting
(\ref{estima1}). Then we
substitute (\ref{potp}) for $V_0$ in (\ref{estima1}) and obtain
\beq
\label{estima6p}
| \pscal{}{\varphi_j(.;k)}{V_0 P_n \varphi_j(.;k)} |
\leq  \mathcal{V}_0
\sum_{m=0}^n | \alpha_m^{(j)}(k) | \left( I_{p,m,-}^{(j)}(k)  +
I_{p,m,+}^{(j)}(k) \right),
\eeq
where
\beq
\label{pIpm}
I_{p,m,\pm}^{(j)}(k) \equiv \int_{\pm x \geq L \slash 2} \left( \pm x - L
\slash 2 \right)^p | \varphi_j(x;k) | | \psi_m(x;k) | dx.
\eeq
We are also left with the task of computing an upper bound for
$I_{p,m,\pm}^{(j)}(k)$. \\

\noindent
{\bf Step 2 : Integral Estimates.}\\
1.\ Let $k$ be in $\omega_j^{-1}(\Delta_n)_-$. By applying Lemma
\ref{lm-WNE} once more we can choose the magnetic strength $B$ sufficiently
large so the quadratic potential
\beq
\label{QuadPot}
Q_m(x;k) \equiv (Bx-k)^2-(2m+1)B
\eeq
is positive in the region $x \geq 0$. Consequently the eigenfunction
$\psi_m(.;k)$ of $h_L(k)$ decays exponentially in the region $x \geq 0$
since this is
an $\Hp{\R}$-solution to the differential equation $\psi''(x) = Q_m(x;k)
\psi(x)$. It follows from this (see Lemma \ref{lm-TE} in Appendix 2) that:
\[ 0 \leq  \mathcal{V}_0 I_{p,m,+}^{(j)}(k) \leq \frac{2}{L} \sqrt{(2n+c)
B},\ m=0,1,\ldots,n. \]
Next combining this estimate with the normalization condition
$\sum_{m=0}^{n} | \alpha_m^{(j)}(k) |^2 \leq 1$, leads to
\beq
\label{estima7p}
0 \leq \mathcal{V}_0
\sum_{m=0}^n | \alpha_m^{(j)}(k) | I_{p,m,+}^{(j)} \leq \frac{2}{L}
\sqrt{(2n+c)n B}.
\eeq
2.\ We turn now to computing an upper bound involving $I_p^{(j)}(k)$ on the
integral
\[ I_{p,m,-}^{(j)}(k) = \int_{-\infty}^{-L \slash 2} (-x-L \slash 2)^p
\varphi_j(x;k) \psi_m(x;k) dx. \]
This can be made by applying the Cauchy-Schwartz inequality
\beq
\label{Ipmm1}
I_{p,m,-}^{(j)}(k) \leq \left( \int_{-\infty}^{-L \slash 2} (-x-L \slash
2)^{p+1} \psi_m(x;k)^2 dx \right)^{\demip} I_p^{(j)}(k)^{\demip},
\eeq
and bounding the prefactor $\int_{-\infty}^{-L \slash 2} (-x-L \slash
2)^{p+1} \psi_m(x;k)^2 dx$ as in Lemma \ref{lm-ipm} in Appendix 2.
Namely, we assume that
\begin{equation}
\label{p-V3}
\mathcal{V}_0  \geq (2n+c) B^{\frac{p+2}{2}},
\end{equation}
so there is a constant $C_m^-(n,p)$ independent of $B$ such that:
\[ 0 \leq \int_{-\infty}^{-L \slash 2} (-x-L \slash 2)^{p+1} \psi_m(x;k)^2
dx \leq C_{m}^-(n,p)^2 B^{-\frac{p+1}{2}}. \]
This, together with (\ref{Ipmm1}) involves
\[ I_{p,m,-}^{(j)}(k) \leq C_{m}^-(n,p) B^{-\frac{p+1}{4}}
I_p^{(j)}(k)^{\demip}, \]
so we get
\beq
\label{p-Ipm}
\sum | \alpha_m^{(j)}(k)  | I_{p,m,-}^{(j)}(k) \leq C^-(n,p)
B^{-\frac{p+1}{4}} I_p^{(j)}(k)^{\demip},
\eeq
from the Cauchy-Schwartz inequality and the normalization condition
$\sum_{m=0}^n | \alpha_m^{(j)}(k) |^2 \leq 1$, the constant $C^-(n,p)$ being
defined as
\[ C^-(n,p) \equiv  \left( \sum_{m=0}^n (C_{m}^-(n,p))^2 \right)^{\demip}.
\]

\noindent
{\bf Step 3 : Estimate on the Main Term}\\
By combining now the estimates (\ref{Sharp-lowerbound}), (\ref{estima6p}),
(\ref{estima7p}) and (\ref{p-Ipm}), we end up getting:
\begin{eqnarray*}
& & \mathcal{V}_0^{\demip} I_p^{(j)}(k)^{\demip} \nonumber \\
&\geq &  \frac{B^{\frac{p+5}{4}}}{C^-(n,p) \mathcal{V}_0^{\demip}} \left(
\frac{(a-1)(3-c)}{2(n+1)} - \frac{2}{L}  \sqrt{(2n+c)n} B^{-\demip} \right).
\end{eqnarray*}
This estimate remains valid as long as (\ref{p-V3}) holds true and $B$ is
sufficiently large.
Whence there is a constant $C(n,p) >0$ independent of $B$ such that
\beq
\label{tracestima2}
\mathcal{V}_0 I_p^{(j)}(k) \geq \frac{C(n,p)}{v} (a-1)^2(3-c)^2
B^{\frac{3}{2}},\ k \in \omega_j^{-1}(\Delta_n),\ j=0,1,\cdots,n,
\eeq
provided $\mathcal{V}_0$ has the following expression:
\beq
\label{p-V4}
\mathcal{V}_0 = v (2n+c)  B^{\frac{p+2}{2}}.
\eeq
Here the coupling constant $v$ is taken in $[1,+\infty)$ so the condition
(\ref{p-V3}) is automatically satisfied with this choice of $\mathcal{V}_0$.

Now it is easy to check the bound (\ref{derdiscur1}) on the derivative
$\omega_j'(k)$ follows from (\ref{pderivative}), (\ref{Ipjnegligible}) and
(\ref{tracestima2}).


\subsection{Perturbation of Edge Currents}
\label{sec-PerEdgCur}
We now consider the perturbation of the edge currents by adding a bounded
impurity Potential $V_1(x,y)$ to $H_0$.
As in section 2.3 of \cite{[HS1]} for unbounded geometries, we prove that
the lower bound on the edge currents is stable with respect to
these perturbations provided $\| V_1 \|_\infty$ is not too large
compared with $B$.
We continue to use the same notation as in \cite{[HS1]}. That is, $\Delta_n
\subset \R$ denotes a closed, bounded
interval with $\Delta_n \subset ( E_n (B) , E_{n+1} (B))$, for some $n
\geq 0$.
We can write the interval $\Delta_n$ as in (\ref{defDelta}):
\beq
\label{per-1}
\Delta_n = [ (2n + a)B, (2n+c)B], ~~\mbox{for} ~1 < a < c < 3.
\eeq
We consider a larger interval $\tilde{\Delta}_n$ containing $\Delta_n$,
and with the same midpoint $E \equiv (2n + (a+c) \slash 2)B$,
and of the form
\beq
\label{per-2}
\tilde{\Delta}_n = [ (2n + \tilde{a})B, (2n+\tilde{c})B], ~~\mbox{for}
~1 < \tilde{a} < a < c < \tilde{c} < 3.
\eeq

\begin{theorem}
\label{thm-per}
Let $\Delta_n$ and $V_0$ be as in Lemma \ref{lm-main}.
Let $V_1 (x,y)$ be a bounded potential
and let $E (\Delta_n)$ be the spectral projection for $H = H_0 + V_1$ and
the interval $\Delta_n$. Let
$\psi \in L^2 (\R^2)$ be a state satisfying $\psi = E (\Delta_n ) \psi$.
Let $\phi \equiv E_0 (\tilde{\Delta}_n ) \psi$ and $\xi \equiv E_0 (
\tilde{\Delta}_n^c ) \psi$,
so that $\psi = \phi + \xi$. Let $\phi$ have an expansion as in
(\ref{decomp1})
with coefficients $\beta_j (k)$ satisfying the condition (\ref{nonsympsi})
of Theorem \ref{thm-cur}, that is:
\[ \exists \gamma >0,\ | \beta_j(k) |^2 \geq (1+\gamma^2) | \beta_j(-k)
|^2,\ k \in \omega_j^{-1}(\Delta_n)_-, j=0,1,\cdots,n. \]
Then, we have,
\beq
\label{per-en1}
- \langle \psi, V_y \psi \rangle \geq
B^{1/2} \left(  \frac{\gamma^2}{2+\gamma^2} C_n ( 3 - \tilde{c})^2
(\tilde{a} - 1)^2 - F_n( \| V_1 \| / B )  \right)
  ~\| \psi \|^2,
\eeq
where $C_n>0$ is the constant defined in Lemma \ref{lm-main} and
\bea
& & F_n(\| V_1 \| \slash B ) \nonumber \\
& = & \left( \frac{2}{ \tilde{c} -\tilde{a} } \right)^{1
\slash 2} \left( \frac{c-a}{2} + \frac{ \|V_1 \|}{B} \right)^{1 \slash 2}
\times \left[ 2 \left( 2n + c + \frac{ \|V_1 \|}{B} \right)^{1 \slash 2}
\right.\nonumber \\
& & \left. + \frac{\gamma^2}{2+\gamma^2} C_n ( 3 - \tilde{c} )^2 (\tilde{a}
- 1)^2 \left( \frac{2}{ \tilde{c} -\tilde{a}} \right)^{3 \slash 2} \left(
\frac{c-a}{2} +
\frac{ \|V_1 \|}{B} \right)^{3 \slash 2} \right]. \label{per-en2}
\eea
If we suppose that $\| V_1 \|_\infty < v_1 B$, then for a fixed level $n$,
if $c-a$ and $v_1$ are
sufficiently small (depending on $\tilde{a}$, $\tilde{c}$, and $n$),
there is a constant $D_n > 0$ so that
for all $B$, we have
\beq
- \langle \psi, V_y \psi \rangle \geq
D_n B^{1/2} \| \psi \| .
\eeq
\end{theorem}
\begin{Myproof}
With reference to the definitions (\ref{per-1}) and (\ref{per-2}),
we write the function $\psi$ as
\beq
\label{per-3}
\psi = E_0 (\tilde{\Delta}_n) \psi + E_0 (\tilde{\Delta}_n^c) \psi
\equiv \phi + \xi .
\eeq
Next we use the selfadjointness of $V_y$ in $\Lp{\R^2}$, to write
\begin{eqnarray}
\pscal{}{\psi}{V_y \psi} & = &
\pscal{}{\phi}{V_y \phi}  \label{per-4}\\
& +& \pscal{}{\psi}{V_y \xi} +
\pscal{}{V_y \xi}{\phi},\nonumber
\end{eqnarray}
whence, by using the Cauchy-Schwartz inequality, we obtain
\begin{equation}
\label{per-5}
-\pscal{}{\psi}{V_y \psi} \geq
-\pscal{}{\phi}{V_y \phi}
- 2 \| V_y \xi \|_{\Lp{\R^2}} \| \psi \|.
\end{equation}

The result follows from Theorem \ref{thm-cur} provided we have a good bound
on $\| \xi \|$ and on
$\| V_y \xi \|$. We recall from section 2.3 in \cite{[HS1]} that
\beq
\label{per-6}
\| \xi \|  \leq  \left( \frac{2}{\tilde{c} - \tilde{a}} \right)
~\left( \frac{c-a}{2} + \frac{\|V_1
\|}{B} \right) \| \psi \|,
\eeq
and
\beq
\label{per-7}
\| V_y \xi \|^2 \leq \langle \xi , H_0 \xi \rangle
\leq  ( (2n + c)B + \| V_1 \| ) ~\| \xi \| ~\| \psi \| .
\eeq
The lower bound on the main term in (\ref{per-5}) follows from the estimate
(\ref{curr4}):
\bea
\label{per-8}
- \langle \phi , V_y \phi \rangle &\geq & \frac{\gamma^2}{2+\gamma^2} C_n (
\tilde{a} -1)^2 (\tilde{c} - 3)^2
~B^{1/2} ~\left( \Sum_{j=0}^n \int_{\omega_j^{-1} ( \tilde{\Delta}_n )}
| \beta_j (k)|^2 ~dk \right) \nonumber \\
& \geq & \frac{\gamma^2}{2+\gamma^2} C_n ( \tilde{a} -1)^2 (\tilde{c} - 3)^2
~B^{1/2} ( \| \psi \|^2 - \| \xi
\|^2 ),
\eea
since
\[ \Sum_{j=0}^n \int_{\omega_j^{-1} ( \tilde{\Delta}_n )} | \beta_j(k)|^2
~dk =
\| \phi \|^2 = \| \psi \|^2 - \| \xi \|^2. \]
Combining this lower bound (\ref{per-8}), with the estimate on $\| \xi \|$
in
(\ref{per-6}), and  $\| V_y \xi \|$ in (\ref{per-7}),
we find (\ref{per-en1}) with the constant (\ref{per-en2}). This completes the proof.
\end{Myproof}



\renewcommand{\thechapter}{\arabic{chapter}}
\renewcommand{\thesection}{\thechapter}

\setcounter{chapter}{3} \setcounter{equation}{0}

\section{Two-Edge Geometries: Spectral Properties and the Mourre Estimate}

We now examine the spectral properties of the Hamiltonian $H = H_0
+ V_1$, for suitable perturbations $V_1$, for two-edge geometries,
paralleling the study in
sections 4 and 5 of \cite{[HS1]} for one-edge geometries. We use
the commutator method of Mourre \cite{[CFKS],[M]}.
For two-edge geometries,  an analysis of the dispersion curves for $H_0$
showed that $\omega_j' (k)$ does not have fixed sign. Consequently, the
local commutator used for the one-edge geometries in section
2.5, does not immediately apply. We first construct an appropriate
conjugate operator $S_\alpha$
for $H_0$ with a general confining potential $V_0(x)$.
By standard arguments \cite{[CFKS]},
this proves the existence of absolutely continuous spectrum of $H_0$ at
energies away from the Landau levels for sufficiently large $B$.
Of course, the spectral properties of $H_0$ can be obtained directly from
the direct integral decomposition (\ref{dirsum1})
and an analysis of the spectrum of
$h_0(k)$ defined by (\ref{fibered1}).
This proves that the spectrum of $H_0$ is everywhere purely absolutely
continuous.
The advantage of the Mourre method, however, is that
we can obtain the stability of the absolutely continuous spectrum
between Landau levels
under two classes of perturbations $V_1$. We prove that
the spectrum of $H$ is purely absolutely continuous
if 1) $V_1 (x,y)$ is periodic with respect to $y$ with sufficiently small
period  or 2) $V_1(x,y)$ has
some decay in $y$-direction. These results
are similar to those of Exner, Joye, and Kovarik \cite{[EJK]}.
We point out that for the more general class of perturbations $V_1$ treated
in sections 4 and 5 of \cite{[HS1]}, such as random potentials,
we do not know the spectral type of the operator $H$.
However, we still know that there are states carrying nontrivial edge
currents. As follows from the work of Ferrari and Macris \cite{[FM],[FM2]},
the existence of edge currents is not tied to the spectral properties of
$H$. Indeed, the cylinder geometry model shows that
the full Hamiltonian may have only pure point spectrum, yet there are
nontrivial edge currents. Hence, the existence of edge currents is not
directly tied to the existence of continuous spectrum. We will discuss this
in more detail in section \ref{sec-cylinder}.

\subsection{The Mourre Inequality for $H_0$}

We construct a conjugate operator for $H_0 = H_L (B) + V_0$,
where the confining potential $V_0$ depends only on $x$, as above.
Let $U_{\alpha}=\expo{i \alpha p_y}$, for $p_y = - i \partial_y$, and
for any $\alpha \in \R$, be the
translation group in the $y$-direction defined by
\beq
\label{tr1}
(U_{\alpha} g)(y)=g(y+\alpha).
\eeq
Since the representation is unitary, the operator $S_\alpha$
defined by
\beq
\label{tr2}
S_{\alpha}= \frac{i}{2} (  U_{\alpha} y - y U_{-\alpha} )
\eeq
is easily seen to be selfadjoint on the domain $D_y$ of the operator
multiplication by $y$, since $U_\alpha$ preserves this domain.

We next compute the commutator $i [H_0, S_{\alpha}]$, $\alpha \in \R$.
The operator $S_\alpha$ commutes with $p_x$ and $V_0$. Since $V_y = p_y -
Bx$, it is easy to check
that
\beq
\label{commpjSa}
[ V_y , S_{\alpha}]= \frac{1}{2} ( U_\alpha - U_{- \alpha}) = i \sin
(\alpha p_y) ,
\eeq
so that
\beq
\label{commHoiSa}
i [ H_0, S_{\alpha} ] = - 2 \sin(\alpha p_y) V_y ,
\eeq
as a quadratic form on $D(H_0) \cap D_y$, or as an operator identity on the
core $C_0^\infty (\R^2)$.
We also need to compute the double commutator $[[H_0, S_\alpha ] ,
S_\alpha]$. By formula (\ref{commHoiSa}), we find that
\beq\label{doublecomm1}
[[H_0, S_\alpha ] ,
S_\alpha] = 2i [ \sin ( \alpha p_y), V_y ] = 0.
\eeq
Consequently, a positive commutator will imply absolutely continuous
spectrum (cf.\ \cite{[CFKS]})
in the range of the corresponding spectral projector as in
Proposition \ref{prop-mopar}.

We first derive a
general expression for $\langle \psi , [H_0,i S_\alpha ] \psi  \rangle$,
for $\psi \in E_0(\Delta_n) \Lp{\R^2} \cap D_y$ and $\alpha  \in \R$.
For any $\psi \in D(H_0) \cap D_y$, it follows from (\ref{commHoiSa})
that
\beq
\label{comm4}
\langle \psi, [H_0,i S_\alpha] \psi  \rangle  =  -2 \int_{\R} \sin(\alpha k)
\langle \hat{\psi}(\cdot;k) , \hat{V}_y \hat{\psi}(\cdot; k) \rangle ~dk,
\eeq
where, as above,
$\hat{u}$ denotes the partial Fourier transform of $u$ with respect to
$y$. We assume that $V_0$ satisfies (\ref{condipot}) and choose $| \Delta_n
| \slash B$ small enough so Lemma \ref{lm-cross} holds true.
Writing $\psi \in E_0(\Delta_n) \Lp{\R^2} \cap D_y$ as in (\ref{decomp1}),
we find that
\begin{equation}
\label{mourre1}
\langle \psi, [H_0,i S_\alpha ] \psi \rangle =
- \Sum_{j=0}^n   \int_{\omega_j^{-1} ( \Delta_n )}
\sin(\alpha k) | \beta_j(k) |^2
\omega_j'(k) ~dk.
\end{equation}
Here we used the Feynman-Hellmann formula and the vanishing of the
cross-terms established in Lemma \ref{lm-cross}.
The potential $V_0$ being an even function, this is still the case for
the $\omega_j$'s (see Lemma \ref{lm-symetrie}),
so (\ref{mourre1}) can be rewritten as
\begin{equation}
\label{mourre2}
\langle \psi, [H_0,i S_\alpha ] \psi \rangle =
- \Sum_{j=0}^n   \int_{\omega_j^{-1} ( \Delta_n )_-}
\sin(\alpha k) ( | \beta_j(k) |^2 + | \beta_j(-k) |^2 )
\omega_j'(k) ~dk.
\end{equation}
In order to prove a Mourre estimate, it is necessary to bound the
right side of (\ref{mourre2}) from below by a positive constant times $\|
\psi
\|^2$. This obviously requires a
lower bound on the derivative $\omega_j'(k)$ of the
dispersion curves.

We now examine the case where $V_0$ is either the parabolic confining
potential (\ref{potQuad}) or the sharp confining potential (\ref{potSharp}).


\subsubsection{The Parabolic Confining Potential Case}

Let $\Delta_n$ be as in (\ref{defDeltaG}):
\[ \Delta_n =[(2n+a)B_g,(2n+c)B_g],\ 1<a<c<3. \]
When the confining potential $V_0$ is defined by (\ref{potQuad}), the
dispersion curves $\omega_j(k)$, $j=0,1,\cdots,n$, are parabolas
with equation given by (\ref{Quad-1}). Whence
\[
\omega_j^{-1}(\Delta_n)_- = (-k_j(c),-k_j(a)) \cup (k_j(a),k_j(c)),
\]
with $k_j(x)= B_g^{\troisdemip} \slash g \sqrt{2(n-j)+x - 1}$, for $x=a,c$
and $j=0,1,\cdots,n$, according to (\ref{Quad-3})-(\ref{Quad-4}).
In view of proving the coming proposition, let us notice that
\beq
\label{kseq}
k_j(c)>k_j(a)>k_{j+1}(c)>k_{j+1}(a)>0,\ j=0,1,\ldots,n-1.
\eeq

\begin{proposition}
\label{prop-mopar}
Let $ | \Delta_n | \slash B$ be small enough so Lemma \ref{lm-cross} holds
true. Then for any $0 < \alpha < \pi \slash k_0(c)$, we have
\beq
\label{mourrepar1}
-i E_0 ( \Delta_n ) [ H_0, S_{\alpha} ] E_0 (\Delta_n) \geq
\frac{2g}{B_g^{\demip}} (a-1)^{\demip} s_{\alpha,n}(a,c) E_0(\Delta_n),
\eeq
with the constant
\beq
\label{sac}
s_{\alpha,n}(a,c) \equiv \min \left( \sin ( \alpha k_n(a) ), \sin ( \alpha
k_0(c) ) \right) >0.
\eeq
\end{proposition}
\begin{Myproof}
By combining (\ref{mourre2}) with the explicit expression (\ref{Quad-1}) of
the $\omega_j(k)$'s and bearing in mind the derivative $\omega_j'$ is an odd
function, we have
\begin{eqnarray*}
& & \langle \psi, [H_0,i S_\alpha ] \psi \rangle \\
& = & - 2 \left( \frac{g}{B_g} \right)^2 \Sum_{j=0}^n
\int_{-k_j(c)}^{-k_j(a)}
k \sin(\alpha k)  (| \beta_j(k) |^2 + | \beta_j(-k) |^2)~dk.
\end{eqnarray*}
Recalling now the definition (\ref{sac}), it is easy to check the function
$k \sin (\alpha k)$ is bounded from below by $k_n(a) s_{\alpha}(a,c)$ in
$\cup_{j=0}^n [-k_j(c),-k_j(a)]$.
This, together with the identity
\[ \Sum_{j=0}^n   \int_{-k_j(c)}^{-k_j(a)} (| \beta_j(k) |^2 + | \beta_j(-k)
|^2)~dk = \| \psi \|^2, \]
proves the positivity of the commutator.
\end{Myproof}

\subsubsection{The Sharp Confining Potential Case}

Let $\Delta_n$ be defined by (\ref{defDelta}):
\[ \Delta_n=[(2n+a)B,(2n+c)B],\ 1<a<c<3. \]
As follows from Lemmas \ref{lm-main} and \ref{lm-WNE} under suitable
conditions on
$B$ and $\mathcal{V}_0$, each set $\omega_j^{-1}(\Delta_n)_-$,
$j=0,1,\ldots,n$, is an interval $[-k_j^+, -k_j^-]$ with :
\[ 0 <\frac{BL}{3} < k_j^- < k_j^+. \]
The $\omega_j$ being written in increasing order,
we have in addition $k_j^{\pm}<k_{j-1}^{\pm}$ for all $j=1,\ldots,n$, so
\[ \cup_{j=0}^{n} \omega_j^{-1}(\Delta_n)_- \subset [-k_0^+,-k_n^-]. \]
In particular the following inequality holds true,
\beq
\label{kseq2}
0<\frac{BL}{3} < k_n^- < k_0^+,
\eeq
allowing us to prove the coming statement.

\begin{proposition}
\label{prop-mosharp}
Assume that $| \Delta_n | \slash B$ is sufficiently small so Lemma
\ref{lm-cross} holds true. Then for any $0 < \alpha < \pi \slash k_0^+$, we
have
\beq
\label{sac2}
s_{\alpha,n} \equiv \min \left( \sin ( \alpha k_n^- ), \sin ( \alpha k_0^+ )
\right) >0,
\eeq
and
\beq
\label{mourresharp1}
-i E_0 ( \Delta_n ) [ H_0, S_{\alpha} ] E_0 (\Delta_n) \geq C_n
(a-1)^2(3-c)^2 B^{\demip} s_{\alpha,n} E_0(\Delta_n),
\eeq
provided $B$ is large enough and $\mathcal{V}_0 \geq 2(2n+c)B$. The constant
$C_n>0$ is defined in Lemma \ref{lm-main} and is independent of $B$ and
$\mathcal{V}_0$.
\end{proposition}
\begin{Myproof}
In light of (\ref{kseq2}) it is easy to check that
\[ -\sin (\alpha k) \geq s_{\alpha,n},\ k \in \omega_j^{-1}(\Delta_n)_-,\
j=0,1,\ldots,n, \]
so the result is an obvious consequence of (\ref{mourre2}) and Lemma
\ref{lm-main}.
\end{Myproof}

\subsection{Perturbation Theory and Spectral Stability}

The benefit of a local positive commutator is its stability under
perturbations. We consider two types of perturbations of $H_0$: 1)
Perturbations periodic
in the $y$-direction, and 2) Perturbations decaying in the $y$-direction.
As we mention below, these conditions on the perturbations are much weaker
than what is required using scattering theoretic methods.
In light of the positive commutator results
(\ref{mourrepar1}) and (\ref{mourresharp1}), we will treat both confining
potentials (\ref{potSharp}) and (\ref{potQuad}) simultaneously, only
referring to the explicit lower bound
for the commutator of $H_0$ when needed.

\subsubsection{Perturbations Periodic in the $y$-Direction}

We first treat perturbations $V_1 (x,y)$ satisfying $V_1(x,y+T)=V_1(x,y)$,
for some $T > 0$. Due to the $y$-periodicity of $V_1$, the main property we
will use in this section is the following basic identity:
\beq
\label{trinv1}
[V_1 ,U_{T}]=0.
\eeq

\begin{proposition}
\label{prop-mourre2}
If the magnetic strength $B$ is taken large enough, there is a constant
$c=c(T)>0$ such that
\[
-i E ( \Delta_n ) [ H_0 + V_1 , S_{T} ] E (\Delta_n) \geq c B^{\demip}
E(\Delta_n)
\]
provided $T$, $| \Delta_n | \slash B$ and $v_1 \equiv \| V_1 \|_{\infty}
\slash B$ are sufficiently small.
\end{proposition}
\begin{Myproof}
Let $\Delta_n \subset \tilde{\Delta}_n$ be defined as in (\ref{per-2}).
We decompose  $\psi \in E(\Delta_n) \Lp{\R^2}$ as in (\ref{per-3})
\[ \psi= E_0 (\tilde{\Delta}_n) \psi + E_0 (\tilde{\Delta}_n^c) \psi
\equiv \phi + \xi, \]
and apply (\ref{trinv1}):
\begin{eqnarray}
\pscal{}{\psi}{[H,i S_{T}] \psi} & = &
\pscal{}{\psi}{[H_0,i S_{T}]\psi} \nonumber\\
& = & \pscal{}{\phi}{[H_0,i S_{T}]\phi} + G(\phi,\xi). \label{courantpert1}
\end{eqnarray}
Here the perturbation term $G(\phi,\xi)$ can be expressed, using the partial
Fourier Transform in the $y$ direction, as
\begin{eqnarray}
G(\phi,\xi) & = & \int_{\R} \sin(T k) \pscal{}{\hat{\xi}(.,k)}{(k-Bx)
\hat{\xi}(.,k)} dk  \nonumber \\
& & + 2 \Pre{\int_{\R} \sin(T k) \pscal{}{\hat{\phi}(.,k)}{(k-Bx)
\hat{\xi}(.,k)} dk}. \nonumber
\end{eqnarray}
It is obviously bounded by $2 \| (p_y-Bx) \xi \| \| \psi \|$, whence
\beq
-\pscal{}{\psi}{[H,i S_{T}]\psi} \nonumber
\geq  -\pscal{}{\phi}{[H_0,i S_{T}]\phi} - 2 \| (p_y-Bx) \xi \| \| \psi
\|, \label{courantpert2}
\eeq
according to (\ref{courantpert1}). The main term $(-\pscal{}{\phi}{[H_0,i
S_{T}]\phi})$ is treated by Proposition \ref{prop-mosharp}. Namely,
for sufficiently small $T$ (and under suitable assumptions on $B$ and
$\mathcal{V}_0$) there is a constant $C_n(T) >0$ independent of $B$ and
$\mathcal{V}_0$ such that
\beq
\label{mainterm}
-\pscal{}{\phi}{[H_0,i S_{T}]\phi} \geq C_n(T) (\tilde{a}-1)^2
(3-\tilde{c})^2 B^{\demip} \| \phi \|^2.
\eeq
Recalling (\ref{courantpert2}), it remains to bound
$\| (p_y-Bx) \xi \|$ and $\| \xi \|$ in a convenient way. We shall use the
two following
estimates :
\beq
\label{majo1}
\| \xi \| \leq \frac{c-a+2v_1}{\tilde{c}-\tilde{a}} \| \psi \|,
\eeq
and
\beq
\label{majo2}
\| (p_y-Bx) \xi \| \leq  \left(
\frac{(c-a+2v_1)(2n+c+v_1)}{\tilde{c}-\tilde{a}} \right)^{\demip} B^{\demip}
\| \psi \|,
\eeq
whose proofs are postponed to the end of the demonstration.

Indeed, by  combining inequalities (\ref{courantpert2})-(\ref{mainterm})
with (\ref{majo1})-(\ref{majo2}), we obtain
\begin{eqnarray}\label{courantpert3}
-\pscal{}{\psi}{[H,i S_{T}]\psi} & \geq &
\left[ C_n(T) (\tilde{a}-1)^2 (3-\tilde{c})^2 \left( 1 - \left(
\frac{c-a+2v_1}{\tilde{c}-\tilde{a}} \right)^2 \right) \right. \nonumber
\\
& & \left. - 2 (2n+c+v_1)^{\demip} \left(
\frac{c-a+2v_1}{\tilde{c}-\tilde{a}} \right)^{\demip} \right] B^{\demip} \|
\psi \|^2. \nonumber
\end{eqnarray}
It is clear now the prefactor of $B^{\demip}$ in the righthand side of
(\ref{courantpert3}) can be made positive by taking $c-a$ and $v_1$
sufficiently small relative to
the difference $\tilde{c}-\tilde{a}$.

We turn now to proving (\ref{majo1})-(\ref{majo2}). First, $E$ denoting the
midpoint of $\Delta_n \subset \tilde{\Delta}_n$, we notice that
$(H_0-E)^{-1} \xi$ is well defined, so we have
\begin{eqnarray*}
\| \xi \|^2 & = & \pscal{}{\psi}{\xi}\\
& = & \pscal{}{(H_0-E) \psi}{(H_0-E)^{-1} \xi}\\
& \leq & \| (H-E-V_1) \psi \| \| (H_0-E)^{-1} \xi \|,
\end{eqnarray*}
from the Cauchy-Schwarz inequality. This, together with the two following
basic estimates,
\beq
\label{ineg-per1}
\| (H-E-V_1) \psi \| \leq \left( \frac{| \Delta_n | }{ 2} + \| V_1
\|_{\infty} \right) \| \psi \|,
\eeq
and
\[ \| (H_0-E)^{-1} \xi \| \leq \mbox{dist}^{-1}(E,\tilde{\Delta}_n^c) \| \xi
\|, \]
proves (\ref{majo1}). To show (\ref{majo2}), we combine the obvious
inequality
\[ \| (p_y-Bx) \xi \|^2 \leq \pscal{}{H_0 \xi}{\xi}, \]
with the following identity
\[ \pscal{}{H_0\xi}{\xi}=\pscal{}{H_0 \psi }{\xi} = \pscal{}{(H-V_1)
\psi}{\xi}, \]
then we use the Cauchy-Schwarz inequality once more, getting:
\[ \| (p_y-Bx) \xi \|^2  \leq \| H - V_1 \|) \| \psi \| \| \xi \|. \]
Now (\ref{majo2}) follows from this, (\ref{majo1}) and (\ref{ineg-per1}).
\end{Myproof}


\subsubsection{Perturbations Decaying in the $y$-Direction}

We now consider an impurity potential $V_1=V_1(x,y) \in L^{\infty}(\R^2)$
having ``good'' decay properties in the $y$-direction. More precisely, we
assume that $V_1$ decays fast enough in the $y$-direction so $y V_1(x,y)$
remains bounded in $\R^2$:
\beq
\label{assumeVone}
\| y V_1 \|_{\infty} < \infty.
\eeq
The reason for this additional assumption is the identity,
\begin{eqnarray*}
& & 2 [ V_1, i S_{\alpha} ] \\
& = & ( V_1(x,y+\alpha) - V_1(y) ) U_{\alpha} y - (  V_1(x,y-\alpha) -
V_1(x,y) ) y U_{-\alpha},
\end{eqnarray*}
obtained by a straightforward computation. This entails
\[ | \langle \psi , [ V_1 , i S_{\alpha} ] \psi \rangle | \leq (2 \| y V_1
\|_{\infty} + | \alpha | \| V_1 \|_{\infty} ) \| \psi \|^2,\ \psi \in D(H_0)
\cap D_y,\ \alpha \in \R, \]
which, combined with the proof of Proposition \ref{prop-mourre2}, entails:

\begin{proposition}
\label{prop-mourre3}
Let $B$ be large. Then there is a constant $c=c(\alpha)>0$ such that
\[
-i E ( \Delta_n ) [ H_0 + V_1 , S_{\alpha} ] E (\Delta_n) \geq c B^{\demip}
E(\Delta_n),
\]
provided $\alpha$, $| \Delta_n | \slash B$, $\| V_1 \|_{\infty} \slash B$
and $\| y V_1 \|_{\infty} \slash B^{\demip}$ are sufficiently small.
\end{proposition}


\subsubsection{Remark on the Stability of the Absolutely Continuous
Spectrum for Strips}

Following the idea developed by Macris,
Martin and Pul\'e in \cite{[MMP]} for the half-plane geometry,
we can actually prove $H_0+V_1$ has purely
absolutely continuous spectrum for the two-edge geometry
if the perturbation $V_1$ is bounded and integrable in $\R^2$.
This class of perturbations is weaker than the classes considered above for
which we proved the existence of absolutely continuous spectrum away from
the Landau levels since, roughly speaking, the $L^1$-condition requires
decay in all directions.
The proof of this result
relies on the diamagnetic inequality (see \cite{[CFKS]}, \cite{[Simon3]}):
\beq
\label{diamagnetic}
| \expo{-t H_L(B)} u | \leq \expo{t \Delta} | u |,\ u \in \Lp{\R},\ t \in
\R_+.
\eeq
Here $(-\Delta)$ denotes the nonnegative Laplacian in $\R^2$ and
(\ref{diamagnetic}) holds true for all $B$.
As the confining potential
$V_0$ is nonnegative in $\R^2$,
Kato's inequality (\ref{diamagnetic}) still holds by substituting
$H_0$ for $H_L(B)$, giving
\beq
\label{diamagnetic2}
| \expo{-t H_0} u | \leq \expo{t \Delta}
| u |\ \mbox{and}\ | \expo{-t H} u | \leq \expo{t \| V_1 \|_{\infty}}
\expo{t \Delta} | u |,\ u \in \Lp{\R},\ t \in \R_+,
\eeq
since $V_1$ is bounded.
It follows by explicit calculation
that
$| V_1 |^{\demip}
\expo{t \Delta}$ belongs to the Schmidt class $\mathcal{B}_2(\Lp{\R^2})$
so that
the same is true for  $| V_1 |^{\demip} \expo{-t H_0}$
and $| V_1 |^{\demip} \expo{-t H}$ by (\ref{diamagnetic2}),
with the following estimates:
\beq
\label{diamagnetic3}
\| | V_1 |^{\demip} \expo{-t H_0} \|_{\mathcal{B}_2(\Lp{\R^2})} =  \frac{\|
V_1 \|_1}{\sqrt{2 \pi t}}\ \mbox{and}\ \| | V_1 |^{\demip} \expo{-t H}
\|_{\mathcal{B}_2(\Lp{\R^2})} =  \expo{t \| V_1 \|_{\infty}}
\frac{\| V_1 \|_1}{\sqrt{2 \pi t}}.
\eeq
Let $\mathcal{B}_1(\Lp{\R^2})$ denote the trace class.
To estimate the trace norm of $\expo{-tH} - \expo{-t H_0}$, we use
Duhamel's formula
\beq\label{duhamel1}
\expo{-tH} = \expo{-t H_0} - \int_0^t \expo{sH} V_1 \expo{-s H_0} ds.
\eeq
Due to the estimates (\ref{diamagnetic3}), the H\"older inequality for the
trace norm, and (\ref{duhamel1}), we obtain
\bea\label{traceest1}
\| \expo{-tH} - \expo{-t H_0} \|_{\mathcal{B}_1(\Lp{\R^2})} &\leq&
\int_0^t \| \expo{(s-t)H} V_1 \expo{-s H_0} \|_{\mathcal{B}_1(\Lp{\R^2})}
\nonumber \\
& \leq  & \frac{\| V_1 \|_1^2 \expo{t \| V_1 \|_{\infty}}}{2 \pi}
   \int_0^t \frac{ds}{\sqrt{s(t-s)}} < \infty.
\eea
Whence $\expo{-tH} - \expo{-t H_0}$ is a trace class operator for
all $t>0$ so $H$ has an absolutely continuous spectrum
by the Kato-Rosenblum Theorem and the fact that $H_0$ has purely absolutely
continuous spectrum.


\renewcommand{\thechapter}{\arabic{chapter}}
\renewcommand{\thesection}{\thechapter}

\setcounter{chapter}{4} \setcounter{equation}{0}

\section{Bounded, Two-Edge, Cylindrical Geometry}
\label{sec-cylinder}
We address now the case of a quantum device with bounded cylindrical
geometry.
More precisely, the charged particle is assumed to be moving on
the cylinder $C_D$ of circumference $D>0$ and confined
along the cylinder axis by two boundaries separated by the distance $L>0$.
We define the infinite cylinder as $C_D= \R \times J = \{ (x,y) ~| ~ x \in
\R, y \in J \}$,
where $J$ is an interval with length $D$,
\[ J= [- D \slash 2, D \slash 2 ], \]
and identify $y = - D/2$ with $y = D/2$.
The trajectories of the particle will be bounded in the $x$-direction by
confining potentials.

Let us give now a precise statement of the model.
The Landau Hamiltonian $H_L(B)=p_x^2+(p_y-Bx)^2$ is endowed with
$y$-periodic
boundary conditions
\beq
\label{perio}
\varphi(x,-D \slash 2)=\varphi(x, D \slash 2)\ \mbox{and}\
\partial_y \varphi(x,- D \slash 2)=\partial_y \varphi(x, D \slash 2),
\eeq
making it selfadjoint in $\Lp{C_D}$.
As in the preceding sections, the quantum particle is confined in
the $x$-direction to the strip $[-L \slash 2, L \slash 2]$
by adding to $H_L(B)$ a confining
potential $V_0=V_0(x)$ fulfilling the condition (\ref{StripConfPot}),
\[ V_0(x) \chi_{ \{ |x| < L \slash 2 \} }(x) = 0, \]
and condition (2.7).
The spectrum of $H_0=H_L(B)+V_0$
consists of eigenvalues for energies below $C$, where $C$ is the limit at
infinity of the confining potential as in (2.7).
It follows from Lemma \ref{discretespe}
that the entire spectrum of $H_0$ is discrete in the
case of the power function confining potential (\ref{potp}).
Despite this,
we shall prove that suitable states
$\varphi = E_0(\Delta_n) \varphi$, $\Delta_n \subset (E_n(B), E_{n+1}(B))$,
carry a current of size $B^{\demip}$,
and that this current survives in presence of a sufficiently small
perturbation.
Thus, the existence of the edge current is independent of the spectral type
of the operator.

This result is in accordance with (and complements) the one obtained by
Ferrari and Macris, who have extensively investigated this model
(\cite{[FM]}, \cite{[FM2]}, \cite{[FM3]}, \cite{[FM4]}) in the particular
case where $D=L$. They consider an Anderson-type random potential
$V_{\omega}$ and
prove with large probability (under a rather technical assumption on the
spectra of the Hamiltonians $H_0^{(l)}$ and $H_0^{(r)}$ obtained
respectively by removing the left or the right wall from $H_0$) that the
spectrum of the random Hamiltonian $H_{\omega}=H_0+V_{\omega}$ in an energy
interval $(B+\| V_{\omega} \|_{\infty}, 3 B - \| V_{\omega} \|_{\infty})$
consists in the union of two sets $\sigma_l$ and $\sigma_r$. The eigenvalues
in $\sigma_{\alpha}$, $\alpha=l,r$, are actually small perturbations of
eigenvalues $E_j^{(l)}$ of the half-plane Hamiltonian $H_{0}^{(\alpha)} +
V_{\omega}$ and they show the edge current carried by an associated
eigenstate $\varphi_j^{(\alpha)}$ is of size $D$ (with opposite signs
depending on whether $\alpha=l$ or $r$). Their analysis extends to the case
where $L$ is at least of size $\log D$.

The remaining of this section is organized as follows. After arguing
$\sigma(H_0)$ is pure point, we estimate the current carried by an
eigenstate of $H_0$ in the case of the sharp confining potential
(\ref{potSharp}). Then we extend this estimate to the case of a convenient
wave packet
$\varphi=E_0(\Delta_n) \varphi$ for $\Delta_n \subset (E_n(B), E_{n+1}(B))$
and in presence of a perturbation $V$ sufficiently small relative to $B$.
We point out that the estimates on the edge currents given in the remaining
of this section are obtained unconditionally on the size of $L$ and $B$ and
they hold for general wave packets with energy in between two consecutive
Landau levels.

\subsection{Nature of the Spectrum of $H_L(B)$ and $H_0$}
\label{sec-spectrum}
\subsubsection{The spectrum of $H_L(B)$}
Let us define the Fourier transform $\Fo$ as
$\Fo \varphi (x) = \left( \hat{\varphi}_p(x) \right)_{p \in \Z}$,
where
\begin{equation}
\label{hatphi}
\hat{\varphi}_p(x)=\int_{J} \varphi(x,y)
\frac{\expo{-i k_p y}}{\sqrt{D}} dy\ \mbox{and}\ k_p=\frac{2 \pi}{D}p,
\end{equation}
for any $p \in \Z$ and a.e. $x \in \R$. It is unitary from $\Lp{C_D}$
endowed with the usual
scalar product onto $l^2(\Z;\Lp{\R})$.
Due to the periodic boundary conditions (\ref{perio}), it is standard result
that
\beq
\label{directsum1}
\Fo H_L(B) {\Fo}^* = \sum_{p \in \Z}^{\oplus} h_L(k_p),
\eeq
where $h_L(k)$, $ k \in \R$, still denotes the operator $p_x^2 + (k-Bx)^2$
in $\Lp{\R}$.
The spectrum of
$h_L(k)$ is discrete and does not depend on $k$, $\sigma(h_L(k))=(2 \N +
1)B$, each eigenvalue being simple. For any $m \geq 0$,
the normalized eigenvector $\psi_m(.;k)$ of $h_L(k)$ associated to the
eigenvalue $(2m+1)B$ is given by (\ref{psim}).
By setting
\beq
\label{defEVL}
\Psi_m^{(p)}(x,y) \equiv \psi_m(x;k_p) \frac{\expo{i k_p y}}{\sqrt{D}},\ m
\in \N,\ p \in \Z,
\eeq
we see from (\ref{directsum1}) the set $\{ \Psi_m^{(p)},\ m \in \N,\ p \in
\Z \}$ is an orthonormal basis of $\Lp{C_D}$ which diagonalizes $H_L(B)$, in
the sense that
\beq
\label{specdecomL}
H_L(B)=\sum_{m \geq 0} (2m+1) B \left( \sum_{p \in \Z} | \Psi_m^{(p)}
\rangle \langle \Psi_m^{(p)} | \right).
\eeq
The spectrum of $H_L(B)$ is also purely punctual with $\sigma(H_L)=  (2 \N +
1)B$,
each eigenvalue having infinite multiplicity.

We turn now to describing the spectrum of $H_0=H_L(B)+V_0$.

\subsubsection{Spectrum of $H_0$}
The confining potential $V_0$ being a function of $x$ alone we deduce from
(\ref{directsum1}) that
\beq
\label{directsum2}
\Fo H_0 {\Fo}^* = \sum_{p \in \Z}^{\oplus} h_0(k_p),
\eeq
where $h_0(k)$ is still defined by (\ref{unpert2}). Moreover the effective
potential $V_{eff}(x;k)=(Bx-k)^2+V_0(x)$ is unbounded at infinity so the
resolvent of $h_0(k)$ is compact. We
recall the eigenvalues of $h_0(k)$ are denoted $\omega_m(k)$, $m \in \N$,
the corresponding normalized
eigenfunction being called $\varphi_m (x ; k)$. By setting analogously to
(\ref{defEVL})
\beq
\label{defEVO}
\Phi_m^{(p)}(x,y) \equiv \varphi_m(x;k_p) \frac{\expo{i k_p y}}{\sqrt{D}},\
m \in \N,\ p \in \Z,
\eeq
we obtain in the same way as before that $\{ \Phi_m^{(p)},\ m \in \N,\ p \in
\Z \}$ is an orthonormal basis of $\Lp{C_D}$, and deduce from
(\ref{directsum2}) that
\beq
\label{specdecomO}
H_0  =  \sum_{m \geq 0} \sum_{p \in \Z} \omega_m(k_p) | \Phi_m^{(p)} \rangle
\langle \Phi_m^{(p)} |.
\eeq
This means that $H_0$ has pure point spectrum:
\beq
\label{specH0}
\sigma(H_0) = \{ \omega_m(k_p),\ m \geq 0,\ p \in \Z \}.
\eeq
Let us now consider an impurity potential $V_1 \in L^{\infty}(C_D)$ such
that:
\beq
\label{compact}
V_1(x,y) \chi_{ \{ | x | > L \slash 2 \} }(x) = 0.
\eeq
The bounded potential $V_1$ also has a compact support, hence it is compact.
Now, one question arising from (\ref{specH0}) is to determine whether the
perturbed Hamiltonian $H=H_0+V_1$ has an eigenvalue.
Since $H$ is obtained from $H_0$ by adding a compact perturbation $V_1$,
standard arguments warrant the answer is positive provided
$\sigma(H_0)$ is discrete. We state in the coming lemma that this is the
case for suitable unbounded confining potentials $V_0$.

\begin{lemma}
\label{discretespe}
The spectrum $\sigma(H_0)$ remains discrete provided $V_0$ is nonnegative
and is such that
\beq
\label{unboundpot}
\lim_{ |x| \rightarrow + \infty} V_0(x)= + \infty.
\eeq
\end{lemma}
\begin{Myproof}
Taking account of (\ref{specH0}), we need to show that each eigenvalue
$E=\omega_{m_0}(k_{p_0})$, $(m_0,p_0) \in \N \times \Z$, is isolated and has
finite multiplicity.

The potential $V_0$ being nonnegative we first notice $\omega_m$ is bounded
from below by $E_m(B)$, so the set
\[ M_E(\epsilon) \equiv \{ m \in \N, \omega_m^{-1}((E-\epsilon, E +
\epsilon) ) \neq \emptyset \}, \]
is finite for any $\epsilon >0$.
Next, $V_0$ satisfying (\ref{unboundpot}) we know from Lemma \ref{lm-asyVP}
in Appendix 1 that
\[ \lim_{|k| \rightarrow + \infty} \omega_m(k) = + \infty, \]
so $\omega_m^{-1}((E-\epsilon, E + \epsilon))$ is bounded for any $m \in
M_E(\epsilon)$.

This indicates that $\{ m \in \N,\ p \in \Z,\ \omega_m(k_p) \in (E-\epsilon,
E+\epsilon) \}$ is necessarily a finite set,
proving the result.
\end{Myproof}

Notice that the power function confining potential (\ref{potp}) fulfills
(\ref{unboundpot}), so
$\sigma_e(H)=\sigma_e(H_0)=\emptyset$ by Lemma \ref{discretespe},
whence  $\sigma(H)$ remains discrete in this case.

Moreover, in the particular case where $V_0$ is the sharp confining
potential (\ref{potSharp}), we can argue in the same way as in
the proof of Lemma \ref{discretespe} that each eigenvalue
$\omega_{m_0}(k_{p_0})$ has finite multiplicity.
However it is not clear that the spectrum of $H_0$ remains discrete. Indeed
as $|p|$ goes to infinity, each $\omega_m(k_p)$, $m \in \N$, goes to $E_m(B)
+ \mathcal{V}_0$ by Lemma \ref{lm-asyVP}, so the eigenvalues lying in a
neighborhood of $E_m(B) + \mathcal{V}_0$ may not be isolated.

\subsection{Edge Currents: the Unperturbed Case}
\label{sub-EdgeCurrUnperFinite}
Let $\Delta_n$ for $n \geq 0$, be defined by (\ref{defDelta}),
\[ \Delta_n=( (2n+a)B, (2n+c)B),\ 1 < a < c < 3,\]
and consider a state $\varphi = E_0(\Delta_n) \varphi$.

We want to estimate the current carried by $\varphi$ along the edges of the
free sample $C_D$ (i.e. when the impurity potential $V_1=0$). It turns out
(see below the estimate
(\ref{current7}) of the current carried by a wave packet) this current is
the weighted sum of the currents carried by all the eigenstates
$\Phi_m^{(p)}$, $(m,p) \in \N \times \Z$, such that
\beq
\label{WP1}
\omega_m(k_p) \in \Delta_n.
\eeq
We therefore start by estimating the current carried by such an eigenstate
$\Phi_{m}^{(p)}$, for appropriate indices $m \in \N$ and $p\in \Z_-$. In a
second step we extend this estimate to the case of the wave packet
$\varphi$.

For simplicity, we assume in the remaining of this section that $V_0$ is the
sharp confining potential
(\ref{potSharp}).

\subsubsection{Current Carried by an Eigenstate}
We consider an eigenfunction $\Phi_{m}^{(p)}$ of $H_0$ for some $(m,p)$ in
$\N \times \Z_-$  satisfying (\ref{WP1}).
The current carried by $\Phi_m^{(p)}$ along the left edge of the cylinder
$C_D$ is defined as the expectation
$\pscal{}{\Phi_m^{(p)}}{V_y \Phi_m^{(p)}}$
of the velocity operator $V_y=p_x-Bx$ in the $y$-direction.
By recalling the formal equality $V_y=\frac{1}{2B} (V_0' + i [ H_0, p_x ])$,
the current immediately decomposes in two terms :
\[ \pscal{}{\Phi_m^{(p)}}{V_y \Phi_m^{(p)}}  = \frac{1}{2B}
\left( \pscal{}{\Phi_m^{(p)}}{ V_0'  \Phi_m^{(p)}} +
\pscal{}{\Phi_m^{(p)}}{[H_0,p _x]  \Phi_m^{(p)}} \right). \]
Following the notations of section \ref{sub-Basic}, the second term
\[ \frac{1}{2B} \pscal{}{\Phi_m^{(p)}}{ [H_0,p _x]  \Phi_m^{(p)}}
=  \frac{1}{2B} \pscal{}{\Phi_m^{(p)}}{[H_0-\omega_{m}(k_p),p _x]
\Phi_m^{(p)}}, \]
vanishes according to the Virial theorem, so we have
\beq
\label{current3}
\pscal{}{\Phi_m^{(p)}}{ V_y \Phi_m^{(p)}}=\omega_{m}'(k_{p}),
\eeq
by the Feynman-Hellmann Formula. In light of (\ref{current3}) the following
result follows immediately from Lemma \ref{lm-main}.
\begin{proposition}
\label{prop-cylcurrent}
Let $V_0$ be the sharp confining potential (\ref{potSharp}) and $\Delta_n$
be defined by (\ref{defDelta}), $| \Delta_n | \slash B$ being sufficiently
small so Lemma \ref{lm-cross} is true.
Then, for any $(m,p) \in \N \times \Z_-$ satisfying (\ref{WP1}), there is a
constant $C_{n}>0$ independent of $B$ and $\mathcal{V}_0$ such that
\[ - \langle \Phi_{m}^{(p)} , V_y \Phi_{m}^{(p)} \rangle \geq C_{n}
B^{\demip},\]
provided $B$ is large enough.
\end{proposition}

\subsubsection{Current Carried by a Wave Packet}

We turn now to estimating the current carried along $C_D$ by a the state
$\varphi = E_0(\Delta_n) \varphi$.
The state $\varphi$ decomposes in the orthonormal basis $\{ \Phi_m^{(p)},\ m
\in \N,\ p \in \Z \}$ as
\beq
\label{phi2}
\varphi(x,y) = \sum_{p \in \Z} \sum_{\tiny \begin{array}{c} 0 \leq m \leq n
\\ \omega_m(k_p) \in \Delta_n \end{array}} \beta_m^{(p)} \Phi_m^{(p)}(x,y),
\eeq
where
\beq
\label{defbetamp}
\beta_m^{(p)} = \pscal{}{\varphi}{\Phi_m^{(p)}}.
\eeq
We suppose that $\mathcal{V}_0$ is sufficiently large, more precisely that
\beq
\label{h1}
\mathcal{V}_0 \geq E_{n}(B) + B,
\eeq
so there are only a finite number of index $p$'s involved in the sum
(\ref{phi2}). Indeed, we know from Lemma \ref{lm-asyVP} that
$\lim_{|k| \rightarrow +\infty} \omega_0(k) = E_0(B) + \mathcal{V}_0$ with
$E_0(B) + \mathcal{V}_0 \geq E_{n+1}(B)$ according to (\ref{h1}). Whence
there is necessarily $p_{n}^* \in \N$ such that
\beq
\label{defPmO}
\omega_0(k_{p_{n}^*}) \in \Delta_n\ \mbox{and}\ \omega_0(k_p) \notin
\Delta_n\ \mbox{for all}\ |p| > p_{n}^*.
\eeq
Since $\omega_{n}(k) > \omega_{0}(k)$ for all $n \geq 1$ and $k \in \R$, we
see that $\omega_n(k_p) \notin \Delta_n$ for any $|p| > p_{n}^*$, so
(\ref{phi2}) finally reduces to:
\beq
\label{phi3}
\varphi(x,y) = \sum_{|p| \leq p_{n}^*} \sum_{\tiny \begin{array}{c} 0 \leq m
\leq n \\ \omega_m(k_p) \in \Delta_n \end{array}} \beta_m^{(p)}
\Phi_m^{(p)}(x,y).
\eeq
Henceforth, the current carried by $\varphi$ along the left edge of the
cylinder has the following expression:
\beq
\pscal{}{\varphi}{ V_y  \varphi}
= \sum_{|p|, |p'| \leq p_{n}^*} \sum_{\tiny \begin{array}{c} 0 \leq m,
m'\leq n \\ \omega_m(k_p) \in \Delta_n \\ \omega_{m'}(k_{p'}) \in \Delta_n
\end{array} } \beta_m^{(p)} \overline{\beta_{m'}^{(p')}}
\pscal{}{\Phi_m^{(p)}}{ v_y  \Phi_{m'}^{(p')}}. \label{current5}
\eeq
Actually the crossed terms $\pscal{}{\Phi_m^{(p)}}{ V_y  \Phi_{m'}^{(p')}}$
in (\ref{current5}) vanish
for $p \neq p'$. This can be seen from the two following basic identities
\[ \begin{array}{c}
        \Fo \Phi_m^{(p)}(x) = \left( \delta(s-p) \varphi_m(x;k_p) \right)_{s
\in \Z},\\
        \Fo \left(  V_y  \Phi_{m'}^{(p')} \right) (x) = \left( \delta(s-p')
(x) (k_{p'}-Bx) (x) \varphi_{m'}(x;k_{p'}) \right)_{s \in \Z},
   \end{array}  \]
and from the unitarity of $\Fo$:
\begin{eqnarray*}
\pscal{}{\Phi_m^{(p)}}{ V_y  \Phi_{m'}^{(p')}}
& = & \delta(p'-p) \pscal{}{\varphi_m(.;k_p)}{ (k_p-Bx)
\varphi_{m'}(.;k_{p})} \\
& = & \delta(p'-p) \pscal{}{\Phi_m^{(p)}}{ V_y  \Phi_{m'}^{(p)}}. 
\end{eqnarray*}
As a consequence, (\ref{current5}) can be rewritten as
\beq
\pscal{}{\varphi}{ V_y  \varphi}
=  \sum_{|p| \leq p_{n}^*} \sum_{\tiny \begin{array}{c} 0 \leq m, m'\leq
n \\ \omega_m(k_p) \in \Delta_n \\ \omega_{m'}(k_{p}) \in \Delta_n
\end{array} } \beta_m^{(p)} \overline{\beta_{m'}^{(p)}}
\pscal{}{\Phi_m^{(p)}}{ V_y  \Phi_{m'}^{(p)}}. 
\eeq
Moreover, taking $|\Delta_n| \slash B$ sufficiently small, we have
$\omega_m^{-1}(\Delta_n) \cap \omega_{m'}^{-1}(\Delta_n) = \emptyset$ for
all $m \neq m'$ according to Lemma \ref{lm-cross}, so end up getting:
\beq
\label{current7}
\pscal{\Lp{C_D}}{\varphi}{ V_y  \varphi}
=  \sum_{|p| \leq p_{n}^*} \sum_{\tiny \begin{array}{c} 0 \leq m \leq n \\
\omega_m(k_p) \in \Delta_n \end{array} } | \beta_m^{(p)} |^2
\pscal{\Lp{C_D}}{\Phi_m^{(p)}}{ V_y  \Phi_{m}^{(p)}}.
\eeq
This shows the current carried by $\varphi$ is the
$|\beta_m^{(p)}|^2$-weighted sum of the current carried by the eigenstates
$\Phi_m^{(p)}$ with energy $\omega_m(k_p)$ in $\Delta_n$. In light of
(\ref{current7}) and Proposition \ref{prop-cylcurrent}
we have obtained the following result:
\begin{proposition}
\label{prop-cylcurrent2}
Let $\Delta_n$ be defined by (\ref{defDelta}) and $V_0$ denote the sharp
confining potential (\ref{potSharp}).
Let $\varphi \in \Lp{C_D}$ satisfy $E_0(\Delta_n) \varphi = \varphi$ and
$p_{n}^*$ be the smallest integer satisfying (\ref{defPmO}), so $\varphi$
has expansion as in (\ref{phi3}).
Assume that $\varphi$ is mostly supported on the set of negative wave
numbers $k_p$, i.e. that there is a constant $\gamma>0$ such that
the coefficients $\beta_m^{(p)}$ defined by (\ref{defbetamp}) satisfy
\beq
\label{LeftLocWP}
   | \beta_m^{(-p)} |^2 \geq (1 + \gamma^2) | \beta_m^{(p)} |^2,
\eeq
for all $m=0,1,\ldots,n$, $p=0,1,\ldots,p_{n}^*$ such that $\omega_m(k_p)
\in \Delta_n$. Then there is a constant $C_n>0$ independent of $B$ and
$\mathcal{V}_0$ such that
\[ -\pscal{}{\varphi}{ V_y  \varphi} \geq
C_n \frac{\gamma^2}{2+\gamma^2} \left( a-1 \right)^2 \left( 3 - c \right)^2
B^{\demip} \| \varphi \|^2, \]
provided $B$ is large enough and $| c - a |$ is sufficiently small.
\end{proposition}
\begin{Myproof}
For any $-p_n^* \leq p \leq 0$ and $0 \leq m \leq n$ such that
$\omega_m(k_p) \in \Delta_n$,
Proposition \ref{prop-cylcurrent} assures us that
\[
-\pscal{}{\Phi_m^{(p)}}{ V_y  \Phi_{m}^{(p)}} \geq C_n  \left( a-1 \right)^2
\left( 3 - c \right)^2 B^{\demip},
\]
so the result follows from (\ref{current7}) and (\ref{LeftLocWP}) by just
mimicking the proof of Theorem \ref{thm-cur}.
\end{Myproof}

\subsection{Perturbation Theory}
As in section \ref{sec-PerEdgCur} for the strip geometries we now consider
the perturbation of the edge currents by adding a bounded impurity potential
$V_1(x,y)$ to $H_0$,
and show the lower bound on the edge currents is stable with respect to
these perturbations provided $\| V_1 \|_\infty$ is not too large compared
with $B$.

We continue to use the same notation as in section \ref{sec-PerEdgCur}. That
is, $\Delta_n \subset \R$ denotes a closed, bounded
interval with $\Delta_n \subset ( E_n (B) , E_{n+1} (B))$, for some $n
\geq 0$.
We write the interval $\Delta_n$ as in (\ref{defDelta}):
\[
\Delta_n = [ (2n + a)B, (2n+c)B], ~~\mbox{for} ~1 < a < c < 3.
\]
We consider the larger interval $\tilde{\Delta}_n$ defined by (\ref{per-2}),
containing $\Delta_n$,
and with the same midpoint $E \equiv (2n + (a+c) \slash 2)B$,
\[
\tilde{\Delta}_n = [ (2n + \tilde{a})B, (2n+\tilde{c})B], ~~\mbox{for}
~1 < \tilde{a} < a < c < \tilde{c} < 3. \]
By recalling Proposition \ref{prop-cylcurrent2} and arguing in the same way
as in the proof of Theorem \ref{thm-per} we obtain the following result.

\begin{theorem}
Let $V_0(x)$ be the sharp confining potential (\ref{potSharp}).
Let $V_1 (x,y)$ be a bounded potential
and let $E (\Delta_n)$ denote the spectral projection for $H = H_0 + V_1$
and
the interval $\Delta_n$. Let
$\psi \in L^2 (C_D)$ be a state satisfying $\psi = E (
\Delta_n ) \psi$. Let $\phi \equiv E_0 (
\tilde{\Delta}_n ) \psi$ and $\xi \equiv E_0 ( \tilde{\Delta}_n^c ) \psi$,
so that $\psi = \phi + \xi$. Let $\phi$ have an expansion as in (\ref{phi3})
with coefficients $\beta_m^{(p)}$ satisfying the condition (\ref{LeftLocWP})
of Proposition \ref{prop-cylcurrent2}, that is:
\[ \exists \gamma >0,\ | \beta_m^{(-p)} |^2 \geq (1+\gamma^2) |
\beta_m^{(p)} |^2, \]
for all $m=0,1,\ldots,n$ and $p=0,1,\ldots,p_n^*$ such that $\omega_m(k_p)
\in \Delta_n$.

Then, we have,
\[
- \langle \psi, V_y \psi \rangle \geq
B^{1/2} \left(  \frac{\gamma^2}{2+\gamma^2} C_n ( 3 - \tilde{c})^2
(\tilde{a} - 1)^2 - F_n( \| V_1 \| / B )  \right)
  ~\| \psi \|^2, \]
where $F_n( \| V_1 \| / B )$ has the same expression as in (\ref{per-en2}).
If we suppose that $\| V_1 \|_\infty < v_0 B$, then for a fixed level $n$,
if $c-a$ and $v_0$ are
sufficiently small (depending on $\tilde{a}$, $\tilde{c}$, and $n$),
there is a constant $D_n > 0$ so that
for all $B$, we have
\[
- \langle \psi, V_y \psi \rangle \geq
D_n B^{1/2} \| \psi \| .
\]
\end{theorem}


\renewcommand{\thechapter}{\arabic{chapter}}
\renewcommand{\thesection}{\thechapter}

\setcounter{chapter}{5} \setcounter{equation}{0}

\section{Appendix 1 : Basic Properties of the Eigenvalues and
Eigenfunctions}

Let $V_0=V_0(x) \in L^2_{loc} (\R)$ be nonnegative.  Then the resolvent of
the operator $h_0(k)=h_L(k)+V_0$ is
compact since the effective potential $V_{eff}(x;k)=(Bx-k)^2+V_0(x)$ is
unbounded as $|x| \rightarrow \infty$, so
the spectrum is discrete with only $\infty$ as an accumulation point.
We write the eigenvalues of $h_0(k)$ in increasing order
and denote them by $\omega_j(k)$, $j \geq 0$. The
normalized eigenfunction associated to $\omega_j(k)$ is
$\varphi_j(x;k)$. We recall from Proposition 7.2 in \cite{[HS1]} that the
eigenvalues $\omega_j(k)$, $j \geq 0$, are simple for all $k \in \R$.

In this Appendix we collect the main properties of the eigenvalues and
eigenfunctions of the operator $h_0(k)$ for an even confining potential
$V_0$.

\subsection{Symmetry Properties}
\begin{lemma}
\label{lm-symetrie}
Let $V_0(x) \in L^2_{loc}(\R)$ be a even confining potential. Then for any
$j \in \N$ and $k \in \R$, the eigenvalues $\omega_j(k)$ and eigenfunctions
$\varphi_j(x;k)$  of
$h_0(k)$ satisfy:
\[ \begin{array}{cl}
             (i) & \omega_j(-k)=\omega_j(k)\\
             (ii) & \varphi_j(-x;-k) = \pm \varphi_j(x;k).
           \end{array}  \]
\end{lemma}
\begin{Myproof}
The operation $P$ that implements $x \rightarrow (-x)$ satisfies $P \dom
h_0(k)= \dom h_0(-k)$ and
$P h_0(k) = h_0(-k) P$. This entails
\beq
\label{eqvp}
h_0(-k) P \varphi_j(x;k) = \omega_j(k) P \varphi_j(x;k),
\eeq
so
$\omega_j(k)$ is an eigenvalue of $h_0(-k)$, and there is necessarily some
$m_k \geq 0$ such that
$\omega_j(k) = \omega_{m_k}(-k)$.\\
Since this is true for any $q \neq k$,
we can find
$m_q \geq 0$ such that $\omega_j(q) = \omega_{m_q}(-q)$.
Moreover $\omega_j$ being a continuous function, $\omega_{m_q}(-q)$
goes to $\omega_{m_k}(-k)$ as $q$ goes to $k$, so
$m_q=m_k$ by the simplicity of the eigenvalues. Therefore
$m_k$
does not depend on $k$.
By writing now $m$ instead of $m_k$ we have shown that
\[ \omega_j(k) = \omega_{m}(-k),\ \forall k \in \R. \]
It follows in particular from this that $\omega_n(0) = \omega_{m}(0)$ so we
immediately get
$m=n$ from the simplicity of the eigenvalues once more.\\
To prove $(ii)$, we substitute $(-k)$ for $k$ in (\ref{eqvp}) and use $(i)$,
getting
\[ h_0(k) \varphi_j(-x;-k) = \omega_j(k) P \varphi_j(-x;-k). \]
Now the result follows from the simplicity of the real valued eigenfunction
$\varphi_j(x;k)$ together with the normalization condition
$\| \varphi_j(.;\pm k) \| = 1$.
\end{Myproof}

\subsection{Asymptotic Behavior and Separation of the Dispersion Curves}
We show below that the asymptotic behavior w.r.t. $k$ of the eigenvalue
$\omega_j(k)$, $j \in \N$, depends on whether the confining potential $V_0$
is bounded at infinity or not. More precisely, we assume $V_0$ satisfies
(\ref{condipot}): There is a generalized constant $0 < C \leq \infty$ such
that
\[ \left\{ \begin{array}{cl}
(a) & 0 \leq V_0(x) \leq C,\ \forall x \in \R\\
(b) & \lim_{|x| \rightarrow \infty} V_0(x) = C.
\end{array} \right.
\]
We now deduce from the assumption (\ref{condipot}) the eigenvalue
$\omega_j(k)$ converges to $E_j(B)+C$ or $+\infty$, depending on whether the
constant $C$ in (\ref{condipot}) is finite or infinite. As a corollary, we
show in Lemma \ref{lm-separated} the dispersion curves remain separated.

\subsubsection{Asymptotic behavior of $\omega_j$}
\begin{lemma}
\label{lm-asyVP}
Let $V_0$ fulfill (\ref{condipot}). Then, for any $j \in \N$, we have:
\bea
(i) & \lim_{| k | \rightarrow + \infty} \omega_j(k) = E_j(B) + C &
\mbox{if}\ 0 < C < \infty \label{asySP}\\
(ii) & \lim_{| k | \rightarrow + \infty} \omega_j(k) = +\infty & \mbox{if}\
C = \infty. \label{asyPP}
\eea
\end{lemma}
\begin{Myproof}
Due to Lemma \ref{lm-symetrie} it is enough to show the result for positive
$k$.\\
{\it Case (i).}
We first deduce from operator inequality $h_0(k) \leq h_L(k) + C$, which
obviously follows
from (\ref{condipot})(a), that
\beq
\label{limit1}
\omega_j(k) \leq E_j(B) + C.
\eeq
We next fix $\varepsilon \in (0,1)$ and derive from (\ref{condipot})(b)
there is necessarily $x_{\varepsilon}>0$ such that
\beq
\label{eqminpot}
V_0(x) \geq C - \varepsilon,\ \forall |x| > x_{\varepsilon}.
\eeq
Let $\varphi$ be a normalized function in the domain of $h_0(k)$. By
combining the following basic inequality
\[ \pscal{\Lp{\R}}{h_L(k) \varphi}{\varphi} \geq (1-\varepsilon)
\pscal{\Lp{\R}}{h_L(k) \varphi}{\varphi} + \varepsilon \int_{|x| \leq
x_{\varepsilon}} (Bx-k)^2 | \varphi(x) |^2 dx, \]
with (\ref{eqminpot}), we have
\bea
\pscal{\Lp{\R}}{h_0(k) \varphi}{\varphi}
& = & \pscal{\Lp{\R}}{h_L(k) \varphi}{\varphi} +  \int_{\R} V_0(x) |
\varphi(x)|^2 dx \nonumber \\
& \geq & (1-\varepsilon) \pscal{\Lp{\R}}{h_L(k) \varphi}{\varphi} + C -
\varepsilon + R_{\varepsilon}, \label{limit2}
\eea
where the remaining term is $R_{\varepsilon} \equiv \int_{|x| \leq
x_{\varepsilon}} \left( \varepsilon (Bx-k)^2 - C \right) | \varphi(x)|^2
dx$.
Since $\varepsilon (Bx-k)^2 - C  \geq 0$ on $[- x_{\varepsilon},
x_{\varepsilon}]$ for all
$k \geq k_{\varepsilon} \equiv B x_{\varepsilon} + \sqrt{C\slash
\varepsilon}$, (\ref{limit2}) immediately leads to
\[ \pscal{\Lp{\R}}{h_0(k) \varphi}{\varphi} \geq (1-\varepsilon)
\pscal{\Lp{\R}}{h_L(k) \varphi}{\varphi} + C - \varepsilon,\ k \geq
k_{\varepsilon}. \]
Let $\mathcal{M}_j$ denote a $j$-dimensional submanifold of $\dom h_0(k)$,
$j=0,1,2,\cdots,n$. It follows from the above inequality and the
Max-Min Principle that
\[ \omega_j(k) \geq \min_{\varphi \in \mathcal{M}_j^{\perp},\ \| \varphi
\|=1}
   \pscal{\Lp{\R}}{h_0(k) \varphi}{\varphi} \geq \min_{\varphi \in
\mathcal{M}_j^{\perp},\ \| \varphi \|=1}
   (1-\varepsilon) \pscal{\Lp{\R}}{h_L(k) \varphi}{\varphi} + C -
\varepsilon, \]
so we obtain
\[ \omega_j(k) \geq (1-\varepsilon) E_j(B) + C - \varepsilon,\ \forall k
\geq k_{\varepsilon}, \]
by taking the $\max$ over the $\mathcal{M}_j$'s.
Now (\ref{asySP}) follows from this and (\ref{limit1}).\\
{\it Case (ii).} The function $V_0$ being nonnegative according to
(\ref{condipot})(a), the effective potential $V_{eff}(x;k)= V_0(x)  +
(Bx-k)^2$ satisfies
\[ V_{eff}(x;k) \geq \tilde{V}(x;k),\ x \in \R,\ k \geq 0, \]
where
\[  \tilde{V}(x;k) = \left\{ \begin{array}{cl}
   k^2 \slash 4 & \mbox{if}\ x \in (-\infty,k \slash (2B)) \cup (3k \slash
(2B), + \infty)\\
   \inf_{x \in [k \slash (2B), 3 k \slash (2B) ]} V_0(x) & \mbox{if}\ x \in
[ k \slash (2B), 3 k \slash (2B) ]. \end{array} \right. \]
Hence $V_{eff}(.;k)$ is uniformly bounded from below by
\[ \tilde{V}(k) \equiv \min( k^2 \slash 4, \inf_{x \in [k \slash (2B), 3 k
\slash (2B) ]} V_0(x)), \]
with $\lim_{k \rightarrow +\infty} \tilde{V}(k) = + \infty$ according to
(\ref{condipot})(b).
This, together with the obvious estimate $h_0(k) \geq \tilde{V}(x;k)$ proves
(\ref{asyPP}).
\end{Myproof}

\subsubsection{Separation of the Dispersion Curves}
\begin{lemma}
\label{lm-separated}
If $V_0$ satisfies (\ref{condipot}), then for all $j \in \N$ we have
\bea
(i) & \hspace*{-1cm} \inf_{k \in \R} \left( \omega_{j+1}(k) - \omega_j(k)
\right) > 0 & \mbox{if}\ 0 < C <+\infty. \label{eq-sep1} \\
(ii) & \forall X >0,\ \inf_{| k | \leq X} \left( \omega_{j+1}(k) -
\omega_j(k) \right) > 0 & \mbox{if}\ C = +\infty. \label{eq-sep1bis}
\eea
\end{lemma}
\begin{Myproof}
The constant $C$ being finite, let us suppose that
\[ \inf_{k \in \R} \left( \omega_{j+1}(k) - \omega_j(k) \right) = 0, \]
for some $j \in \N$. There would also be a sequence $(k_m)_{m \geq 1}$ of
real numbers, such that
\beq
\label{eq-sep2}
0 \leq \omega_{j+1}(k_m) - \omega_j(k_m) < \frac{1}{m},\ m \geq 1.
\eeq
Due to the evenness of $\omega_j$ and $\omega_{j+1}$, the $k_m$ could
actually be chosen nonnegative, and for all $B >0$, we deduce from Lemma
\ref{lm-asyVP} the sequence
$(k_m)_{m \geq 1}$ would be necessarily bounded. Therefore we could build a
subsequence $(k_{m'})_{m'}$ of $(k_m)_m$  that converges to $k^* \in \R_+$.
Hence, by substituting $m'$ for $m$ in
(\ref{eq-sep2}) and taking the limit as $m'$ goes to infinity, we would have
\[ \omega_j(k^*)=\omega_{j+1}(k^*), \]
since $\omega_j$ and $\omega_{j+1}$ are continuous functions. This would
mean $\omega_j(k^*)$ is a doubly-degenerated eigenvalue of $h_0(k^*)$, a
contradiction to the simplicity of the eigenvalues of $h_0(k)$, $k \in \R$.

Evidently the case $C=+\infty$ is obtained by arguing in the same way as
before since the parameters $k$ considered in this case are taken in a
bounded set.
\end{Myproof}



\renewcommand{\thechapter}{\arabic{chapter}}
\renewcommand{\thesection}{\thechapter}

\setcounter{chapter}{6} \setcounter{equation}{0}

\section{Appendix 2 : Technical Estimates for the Power Function Confining
Potential}

We collect in Lemmas \ref{lm-TE} and \ref{lm-ipm} two technical estimates
used in section \ref{sub-p-Estimate} for the calculation of the lower bound
(\ref{derdiscur1}) on the edge current, in the particular case where the
confining potential is the power function (\ref{potp}).

Though Lemma \ref{lm-TE} is actually valid for more general confining
potentials, we assume for simplicity in this appendix that
$V_0$ denotes the power function confining potential (\ref{potp}).

\subsection{Bounding Eigenfunctions in the Classically Forbidden Region}
\begin{lemma}
\label{lm-TE}
Upon taking $B$ sufficiently large, we have
\[ 0 \leq \int_{\R_+} V_0(x) \varphi_j(x;k) \psi_m(x;k) dx \leq  \frac{L}{2}
\sqrt{(2n+c) B}, \]
for all  $k \in \omega_j^{-1}(\Delta_n)_-$ and $j,\ m = 0,1,\ldots,n$.
\end{lemma}
\begin{Myproof}
Let $k$ be in $\omega_j^{-1}(\Delta_n)_-$. We know from Lemma \ref{lm-WNE}
that $k \leq -BL \slash 3$ provided $B$ is sufficiently large, so the
effective potential
$W_j(x;k)$ defined in (\ref{EffPot}) is positive in the region $x \geq 0$.
As a consequence the non-zero $\Hp{\R}$-solution  $\varphi_j(.;k)$ to the
differential equation
\beq
\label{DE1}
\varphi''(x) = W_j(x;k) \varphi(x),
\eeq
does not vanish in $\R_+$, according to Proposition 8.1 in \cite{[HS1]}.
Moreover $\varphi_j(L \slash 2;k)$ being chosen positive, we have in
addition:
\beq
\label{s1d}
\varphi_j'(x;k) < 0\ \mbox{and}\ \varphi_j''(x;k)>0,\ x \geq 0.
\eeq
This, together with inequality $\| \varphi_j'(.;k) \|^2 \leq \omega_j(k)$,
which immediately follows from (\ref{DE1}), involves
\beq
\label{appT1}
\varphi_j'(L \slash 2;k)^2 \leq \frac{2}{L} \int_0^{L \slash 2}
\varphi_j'(x;k)^2 dx \leq \frac{2}{L} (2n+c)B.
\eeq
Similarly, $B$ being taken sufficiently large so the quadratic potential
$Q_m(x;k)$ defined in (\ref{QuadPot}) is positive in the region $x \geq 0$,
the function $\psi_m(x;k)$ may be taken positive in $\R_+$, with
\beq
\label{s2d}
\psi_m'(x;k) < 0,\ x \geq 0,
\eeq
since this is a non-zero $\Hp{\R}$-solution to the differential equation
$\psi''(x) = Q_m(x;k) \psi(x)$.
Using the normalization condition  $\| \psi_m(.;k) \| = 1$, it follows from
this that
\beq
\label{appT2}
\psi_m(L \slash 2;k)^2 \leq \frac{2}{L} \int_0^{L \slash 2} \psi_m(x;k)^2 dx
\leq \frac{2}{L}.
\eeq
We turn now to estimating the integral $\int_{\R_+} V_0(x) \varphi_j(x;k)
\psi_m(x;k) dx$. By reference to equation (\ref{DE1}) we substitute the
expression $-\varphi_j''(x;k) - ( (Bx-k)^2 - \omega_j(k) ) \varphi_j(x;k)$
for
$V_0(x) \varphi_j(x;k)$ in the integrand, getting,
\beq
\label{appT3}
\int_{0}^{+\infty} V_0(x) \varphi_j(x;k) \psi_m(x;k) \leq \int_{L \slash
2}^{+\infty} \varphi_j''(x;k) \psi_m(x;k) dx,
\eeq
since $( (Bx-k)^2 - \omega_j(k) ) \varphi_j(x;k) \psi_m(x;k)$ is nonnegative
in the region $x \geq 0$. An integration by parts in the r.h.s. of
(\ref{appT3}) now provides
\begin{eqnarray*}
& & \int_{0}^{+\infty} V_0(x) \varphi_j(x;k) \psi_m(x;k) \\
& \leq  & \left| \varphi_j'(L \slash 2;k) \right| \psi_m(L \slash 2;k) -
\int_{L \slash 2}^{+\infty} \varphi_j'(x;k) \psi_m'(x;k) dx,
\end{eqnarray*}
the last integral being positive according to (\ref{s1d}) and (\ref{s2d}).
The result follows from this together with (\ref{appT1}) and
(\ref{appT2}).
\end{Myproof}

\subsection{Bounding Eigenfunctions Outside the Classically Forbidden
Region}
Bounding the integral $\int_{(-\infty,-L \slash 2)} (-x-L \slash 2)^{p+1}
\psi_m(x;k)^2 dx$ as in Lemma \ref{lm-ipm} requires a slightly different
strategy from the one used in the proof of Lemma \ref{lm-TE}. Indeed, for $k
\in \omega_j^{-1}(\Delta_n)_-$, it is not guaranteed the set $(- \infty, L
\slash 2)$ is entirely in the classically forbidden region of $h_L(k)$ for
the energy $(2m+1)B$.
This can be seen from the fact the quadratic potential $Q_m(x;k)$ defined in
(\ref{QuadPot}) vanishes at the coordinates $x_{\pm}= k \slash B \pm
\sqrt{2m+1} \slash B^{1 \slash 2}$, which, in light of Lemma \ref{lm-WNE},
may belong to $(- \infty, L \slash 2)$.

In view of Lemma \ref{lm-ipm} (in the particular case where $V_0$ is the
power function confining potential (\ref{potp})) we actually need a more
precise bound from below on the set $\omega_j^{-1}(\Delta_n)_-$, than the
one given by Lemma \ref{lm-WNE}. This is the purpose of the coming Lemma.

\subsubsection{Wave Numbers Estimate Revisited}
\begin{lemma}
\label{lm-p1}
Any given $\delta >0$ we have
\[ \omega_j^{-1}(\Delta_n)_- \subset  [-B(L \slash 2 + \delta) -
\sqrt{(2n+c)B},0],\ j=0,1,\cdots,n,\]
provided
\beq
\label{p-V2}
\mathcal{V}_0 \geq (2n+c)B \slash \delta^p.
\eeq
\end{lemma}
\begin{Myproof}
The estimation on the upper bound of $\omega_j^{-1}(\Delta_n)_-$ following
immediately from its definition, it only remains to prove the estimate on
the lower bound.
Actually the eigenvalues $\omega_j(k)$, $j=0,1,\cdots,n$, of $h_0(k)$, being
written in ascending order, it is enough to prove the result for $j=0$. To
do that, we consider
a normalized function $\varphi$ in the domain of $h_0(k)$, $k \in \R$, and
apply the definition (\ref{potp}) of $V_0$. We obtain:
\bea
& & \pscal{}{h_0(k) \varphi}{\varphi} \nonumber \\
& = & \pscal{}{h_L(k) \varphi}{\varphi} + \mathcal{V}_0 \int_{L \slash
2}^{+\infty} (x - L \slash 2)^p  (| \varphi(x) |^2 + | \varphi(-x) |^2) dx
\nonumber \\
& \geq & \pscal{}{h_L(k) \varphi}{\varphi} + \mathcal{V}_0 \int_{L \slash
2+\delta}^{+\infty} (x - L \slash 2)^p  (|\varphi(x) |^2 + | \varphi(-x)
|^2) dx \nonumber\\
& \geq & \pscal{}{h_L(k) \varphi}{\varphi} + \mathcal{V}_0 \delta^p
\int_{|x| \geq L \slash 2+\delta} | \varphi(x) |^2  dx, \label{minmax0}
\eea
for any $\delta>0$. Using the normalization condition $\| \varphi \| = 1$
together with the obvious operators comparison $h_L(k) \geq (Bx-k)^2$, we
deduce from (\ref{minmax0}) that
\bea
& & \pscal{}{h_0(k) \varphi}{\varphi} \nonumber \\
& \geq & \mathcal{V}_0 \delta^p + \int_{|x| <L \slash 2+\delta}  ((Bx-k)^2 -
\mathcal{V}_0 \delta^p) |\varphi(x) |^2 dx. \label{minmax1}
\eea
Let us assume now that $k \leq k_{\delta} \equiv -B \left( L \slash 2
+\delta
\right) - \sqrt{\mathcal{V}_0 \delta^p}$ so
$\omega_0(k) \geq \mathcal{V}_0 \delta^p$ for all $k \leq k_{\delta}$ from
the Min-Max Principle. This means that
\beq
\label{minmax3}
\inf \omega_0^{-1}(\Delta_n)_- \geq -B \left( L \slash 2 + \delta \right) +
\sqrt{(2n+c)B},
\eeq
in the particular case where $\mathcal{V}_0=(2n+c) B \slash \delta^p$. To
achieve the proof it is enough to notice that (\ref{minmax3}) remains valid
for $\mathcal{V}_0 > (2n+c)B \slash \delta^p$ since
$\omega_0(k)$ is an increasing function of $\mathcal{V}_0$.
\end{Myproof}

Armed with this Lemma we can prove the main result of this section.

\subsubsection{The Main Result}

\begin{lemma}
\label{lm-ipm}
There is a constant $C_m(n,p)>0$ independent of $B$ and $\mathcal{V}_0$ such
that,
\beq
\label{eq-ipm0}
\int_{-\infty}^{-L \slash 2} (-x-L \slash 2)^{p+1} \psi_m(x;k)^2 dx \leq
C_{m}(n,p) B^{-\frac{p+1}{2}},\ k \in \omega_j^{-1}(\Delta_n)_-,
\eeq
provided
\beq
\label{condpotp}
\mathcal{V}_0  \geq (2n+c) B^{\frac{p+2}{2}}.
\eeq
\end{lemma}
\begin{Myproof}
Let us define the constant
\[ h_m \equiv \sup_{u \in \R} H_m(u) \expo{-u^2 \slash 4}, \]
where $H_m$ still denotes the $m^{\tiny \mbox{th}}$ Hermite polynomial
function. The main ingredient of the proof is the following estimate
\beq
\label{boundpsim2}
| \psi_m(x;k) | \leq \left( \frac{B}{\pi} \right)^{\quartp}
\frac{h_m}{\sqrt{2^m m!}} \expo{-B \slash 4 (x-k \slash B)^2},\ k \in
\omega_j^{-1}(\Delta_n)_-,
\eeq
which obviously follows from the explicit expression (\ref{psim}) of
$\psi_m(x;k)$. Indeed, by substituting the r.h.s. of (\ref{boundpsim2}) for
$\psi_m(x;k)$ in the integral in
(\ref{eq-ipm0}), we obtain
\bea
& & \int_{-\infty}^{-L \slash 2} (-x-L \slash 2)^{p+1} \psi_m(x;k)^2 dx
\nonumber\\
& \leq & \left( \frac{B}{\pi} \right)^{\demip} \frac{h_m^2}{2^m m!} \int_{L
\slash 2}^{+\infty} (x-L \slash 2)^{p+1} \expo{-B \slash 2 (x+k \slash
B)^2}dx, \label{eqipm1}
\eea
so we are left with the task of bounding the preceding integral, called
$J_p$ in the remaining of this proof. To do that, we study two cases
separately.\\
{\it First Case: $k \slash B \geq - L \slash 2$}. In this case, it is enough
to notice that $x+ k \slash B \geq x - L \slash 2 \geq 0$ for all $x \geq L
\slash 2$, and
use the change of variable $t=\sqrt{B} ( x - L \slash 2 )$, getting
\begin{eqnarray}
J_p & \leq  & \int_{L \slash 2}^{+\infty} (x-L \slash 2)^{p+1} \expo{-B
\slash 2 (x-L \slash 2)^2}dx \nonumber\\
& \leq &  \left( \int_0^{+\infty} t^{p+1}  \expo{-t^2 \slash 2} dt \right)
B^{-\frac{p+2}{2}},\label{estima10p}
\end{eqnarray}
so (\ref{eq-ipm0}) immediately follows from this and from (\ref{eqipm1}).\\
{\it Second Case: $k \slash B < - L \slash 2$}.
Let us decompose the integral $J_p$ into two terms :
\begin{eqnarray}
J_p & = & \int_{L \slash 2}^{-(L \slash 2 + 2 k \slash B)} (x-L \slash
2)^{p+1} \expo{-B \slash 2 (x+k \slash B)^2}dx \nonumber\\
& + & \int_{-(L \slash 2 + 2 k \slash B)}^{+\infty} (x-L \slash 2)^{p+1}
\expo{-B \slash 2 (x+k \slash B)^2}dx. \label{estima11p}
\end{eqnarray}
The first integral can be treated by applying Lemma \ref{lm-p1} for
$\delta=B^{-\demip}$. We get that
\beq
\label{eq-imp1}
0 \leq x - L \slash 2 \leq -2( L \slash 2 + k \slash B) \leq 2 B^{-\demip} (
1 + \sqrt{2n+c}),
\eeq
provided (\ref{condpotp}) is satisfied, this last condition being obtained
by simply rewriting (\ref{p-V2}) with $\delta=B^{-\demip}$.
This, together with the change of variable $t=\sqrt{B} ( x + k \slash B )$
involves:
\bea
& & \int_{L \slash 2}^{-(L \slash 2 + 2 k \slash B)} (x-L \slash 2)^{p+1}
\expo{-B \slash 2 (x+k \slash B)^2}dx \nonumber\\
& \leq &  2^{p+1} ( 1+ \sqrt{2n+c})^{p+1} \sqrt{\pi} B^{-(p+2) \slash 2}.
\label{estima12p}
\eea
The bound on the second term in (\ref{estima11p}) is obtained by noticing
that
\[ 0 < x - L \slash 2 \leq 2 (x + k \slash B),\ \mbox{for all}\ x \geq -(L
\slash 2 + 2 k \slash B), \]
and using the change of variable $t=\sqrt{B}(x+k \slash B)$ once more:
\begin{eqnarray*}
& & \int_{-(L \slash 2 + 2 k \slash B)}^{+\infty} (x-L \slash 2)^{p+1}
\expo{-B \slash 2 (x+k \slash B)^2}dx \\
& \leq & \left( 2^{p+1}  \int_{0}^{+\infty} t^{p+1} \expo{-t^2 \slash 2} dt
\right) B^{-(p+2) \slash 2}. 
\end{eqnarray*}
In light of (\ref{estima10p}), the result now follows from this,
(\ref{eqipm1}), (\ref{estima11p}) and (\ref{estima12p}).
\end{Myproof}


\end{document}